\documentclass[nohyper]{JHEP3}

\usepackage{epsfig}
\usepackage{amsfonts}
\usepackage{amsmath}
\usepackage{amssymb}
\usepackage{amsbsy}
\usepackage{multirow}

\newcommand{\A}[1]{A^{(#1)}}
\newcommand{\C}[1]{C^{(#1)}}
\newcommand{\Gg}[1]{G^{(#1)}}
\newcommand{\CHI}[1]{\chi^{(#1)}}

\newcommand{\PSI}[1]{\Psi^{(#1)}}

\newcommand{\wdg}{{\scriptscriptstyle \wedge}}

\newcommand{\maj}[1]{\bar{#1}}
\newcommand{\dir}[1]{\bar{#1}^{\mathrm{\textit{\tiny{D}}}}}
\newcommand{\transp}[1]{{#1}^{\mathrm{\textit{\tiny{T}}}}}

\def\rmi{{\rm i}}
\def\rmd{{\rm d}}

\def\calc         {{\cal C}}

\def\call         {{\cal L}}

\def\calp         {{\cal P}}

\newsavebox{\uuunit}
\sbox{\uuunit}
    {\setlength{\unitlength}{0.825em}
     \begin{picture}(0.6,0.7)
        \thinlines
        \put(0,0){\line(1,0){0.5}}
        \put(0.15,0){\line(0,1){0.7}}
        \put(0.35,0){\line(0,1){0.8}}
       \multiput(0.3,0.8)(-0.04,-0.02){12}{\rule{0.5pt}{0.5pt}}
     \end {picture}}
\newcommand {\unity}{\mathord{\!\usebox{\uuunit}}}

\def\a{\alpha}

\def\g{\gamma}
\def\G{\Gamma}

\def\D{\Delta}
\def\l{\lambda}

\def\f{\phi}
\def\m{\mu}
\def\n{\nu}

\def\r{\rho}
\def\s{\sigma}

\newcommand{\OSp}{\mathop{\rm {}OSp}}

\title{Pseudo-supersymmetry and a Tale of Alternate Realities}
\author{E.A. Bergshoeff$^1$, J. Hartong$^1$, A. Ploegh$^1$, J. Rosseel$^2$ and D. Van den Bleeken$^2$
\\
$^1$ Centre for Theoretical Physics, University of Groningen,
Nijenborgh
  4,\\
9747 AG Groningen, The Netherlands \\
{\upshape\ttfamily e.a.bergshoeff, j.hartong, a.r.ploegh@rug.nl}
\\ \\
$^2$Instituut voor Theoretische Fysica, KU Leuven, \\
Celestijnenlaan 200D, B-3001 Leuven, Belgium\\
{\upshape\ttfamily jan.rosseel,
dieter.vandenbleeken@fys.kuleuven.be}
\\ }

\abstract{We discuss how all variant 10d and 11d maximal
supergravities, including star supergravities and supergravities
in different signatures, can be obtained as different real slices
of three complex actions. As an application we study the recently
introduced domain-wall/cosmology correspondence in this approach.
We give an example in 9d and 10d where the domain-wall and
corresponding cosmology can be viewed as different real slices of
the same complex solution. We argue how in this case the
pseudo-supersymmetry of the cosmological solutions can be
understood as the invariance under supersymmetry of a variant
supergravity.} \preprint{UG-07-02 \\ KUL-TF-07/08}

\begin{document}
\newpage
\section{Introduction}
In this paper, we will present a complex formulation of 10- and 11
dimensional supergravity theories. One of the reasons to do this
stems from the so-called 'variant supergravities' in 10 and 11
dimensions, whose existence has been discussed first in
\cite{Hull:1998ym,Hull:1998vg,Hull:1998fh}. It was argued that upon
applying T-dualities along time-like directions new supergravities
are found. In particular, time-like T-duality on the usual type IIA
theory does not lead to the usual type IIB theory, but instead leads
to a different theory, called the type IIB* theory. Similarly, the
type IIA* theory is found as the time-like T-dual of the usual type
IIB supergravity. Note that both type II and type II* theories share
the same space-time signature $(1,9)$. A crucial difference between
type II and type II* is that in the *-theories the RR-forms are
ghosts, i.e. they have wrong-sign kinetic terms. Upon applying more
general dualities, one is also led to type II supergravities in
different signatures. Similarly, it was argued that one should also
consider eleven-dimensional supergravity in different signatures.
For instance, it was shown that the type IIA* theory could be
obtained by dimensional reduction over a time-like direction of 11d
supergravity in signature $(2,9)$.

In the first part of this paper we derive the explicit actions and
supersymmetry variations of these variant supergravities. For
earlier work on the construction of these theories in the IIA and
M-theory case, see \cite{Vaula:2002cn,Nishino:1999vk}. We will adopt
a different approach for constructing the actions and furthermore
include the IIB case. The strategy we will follow in obtaining
actions and supersymmetry transformation rules for these
supergravities, is based on the observations made in
\cite{Bergshoeff:2000qu}. There, it was shown that the superalgebras
underlying these variant supergravities correspond to different
parameterizations of the unique real form of the superalgebra
$\OSp(1|32)$. Our work can be viewed as a continuation of
\cite{Bergshoeff:2000qu}, where now we construct the complex field
theory corresponding to the complex algebra presented there. More
precisely, starting from the complex algebra, one can impose
different reality conditions on the generators. Each choice of
reality conditions gives a real superalgebra underlying one of the
variant supergravities in a specific signature. Similarly we will
start from a single complex action and by imposing different reality
conditions obtain the different variant supergravities.

Another motivation to study such a complex formulation of
supergravity is the domain-wall/cosmology correspondence as
introduced in
\cite{Skenderis:2006jq,Skenderis:2006fb,Skenderis:2006rr}. This
correspondence was introduced in the context of so-called fake
supergravity \cite{Freedman:2003ax}, where one studies systems of
gravity coupled to scalar fields $\phi^A$. The Lagrangian
generically takes the following form:
\begin{equation} \label{lagrfakesugra}
S = \int d^d x\, e \Big[R - \frac{1}{2} G_{AB}(\phi)
\partial_\mu \phi^A \partial^\mu \phi^B - \beta V(\phi) \Big]\,,
\end{equation}
where $G_{AB}(\phi)$ is a metric on the target space spanned by
the scalar fields $\phi^A$, $V(\phi)$ is a potential for the
scalars and $\beta$ represents a sign that can be either $+1$ or
$-1$. The domain-wall/cosmology correspondence is based on the
fact that the existence of a domain-wall solution of the system
(\ref{lagrfakesugra}) with $\beta = +1$ automatically implies the
existence of a cosmological solution of the system
(\ref{lagrfakesugra}) with $\beta = -1$. The domain-wall solutions
generically are 'fake supersymmetric'
\cite{Skenderis:2006jq,Skenderis:2006fb,Skenderis:2006rr}. This
implies that one can write the potential $V$ in terms of a real
superpotential $W$. For the one scalar case this relation
schematically looks like
\begin{equation} \label{VintermsofWdw}
V=2\left((W')^2-W^2\right) \,,
\end{equation}
where $W'=\frac{\delta W}{\delta \phi}$. Furthermore, the
domain-walls allow for the existence of a Killing spinor
$\epsilon$ obeying a Killing spinor equation that can be written
in terms of the superpotential $W$ as follows:
\begin{equation} \label{killspinoreqdw}
(D_\m- W\G_\mu)\epsilon=0 \,.
\end{equation}
In case the Lagrangian (\ref{lagrfakesugra}) can be obtained as a
truncation of a supergravity theory the equations
(\ref{VintermsofWdw},\ref{killspinoreqdw}) can be understood as
arising from the structure of the underlying supergravity theory. In
particular, the Killing spinor equation could in that case be
obtained by putting the supersymmetry transformations of the
fermions equal to zero. However, fake supergravity is much more
general and the Lagrangian (\ref{lagrfakesugra}) can be completely
general and does not need to be related to any supergravity theory.
The mapping between domain-walls and cosmologies implies that
cosmologies also obey a property that looks very much like fake
supersymmetry. In this case, it turns out that the cosmology obeys
similar equations (\ref{VintermsofWdw},\ref{killspinoreqdw}) as its
corresponding domain-wall solution, with the caveat that now the
superpotential $W$ is no longer real but is instead purely
imaginary. Redefining $W = i \tilde{W}$, equations
(\ref{VintermsofWdw},\ref{killspinoreqdw}) become
\begin{eqnarray} \label{VintermsofWkillspinoreqcosm}
& &  V  =  - 2\left((\tilde{W}')^2-\tilde{W}^2\right) \,,
\\
& & (D_\m- i \tilde{W}\G_\mu)\epsilon=0 \,.\label{secondpart}
\end{eqnarray}
Note the change of sign in (\ref{VintermsofWkillspinoreqcosm}),
which indeed corresponds to $\beta=-1$ in (\ref{lagrfakesugra}).
The structure (\ref{VintermsofWkillspinoreqcosm},\ref{secondpart})
for cosmological solutions was called 'pseudo-supersymmetry'
\cite{Skenderis:2006jq,Skenderis:2006fb,Skenderis:2006rr}.

\,\!\!From a supergravity point of view, this correspondence is
rather odd. Supersymmetric domain-wall solutions can be found rather
generically in supergravity theories . For supersymmetric
cosmological solutions this is not true. Furthermore, the
correspondence involves a sign change in the potential that spoils
the supersymmetry of the supergravity theory under consideration.
Finally, in fake supergravity theories, one is usually not concerned
with the reality properties of the (Killing) spinors and one works
with arbitrary Dirac spinors. In real supergravity theories, reality
conditions on the spinors have to be imposed in order to account for
the correct number of degrees of freedom. In this respect, one no
longer has the freedom to take $W$ purely imaginary without
upsetting the reality properties of the supersymmetry rules.

A natural question is whether one can give a meaning to
pseudo-supersymmetry in a real supergravity context. The fact that
the corresponding domain-wall and cosmological solutions differ in
the reality properties of the superpotential suggests that, if one
can give an embedding of the correspondence in supergravity, one
should look for theories in which the spinors obey different
reality properties. A priori, it is possible that there are two
different theories in the same signature (namely $(1,9)$) that
mainly differ in the reality properties of the spinors. This can
then account for a difference in reality properties of the
superpotential and for the sign flip in the potential. We present
an example of this is in the type II and type II* theories in
signature $(1,9)$. Starting from a supersymmetric domain-wall in
type IIA,  the corresponding cosmological solution then turns out
to be a supersymmetric solution of the type IIA* theory.
Pseudo-supersymmetry in this context corresponds to supersymmetry
in a star theory.

As we will show in two examples, the formalism of complex actions
allows for a uniform description of certain domain-walls and
cosmologies. The domain-wall and its cosmological counterpart are
different real slices of one single complex solution of the complex
supergravity. Imposing reality conditions distinguishes the two
backgrounds as solutions of the different real theories and the same
applies to their supersymmetry properties.

This paper is structured as follows. In section \ref{Type II} we
show how to construct a complex action for the type II theories.
Here we explain the method of taking real slices and present
several examples. In section \ref{pseudo} we apply the results of
section \ref{Type II} to the domain-wall/cosmology correspondence.
We present our conclusions in section \ref{disc}. Our conventions
are summarized in appendix \ref{Conventions}. Appendix
\ref{spinors} discusses in detail the reality conditions of the
different spinors appearing in the main text. We have placed a
rather technical discussion about reality conditions for the
vielbeine in appendix \ref{vielbeine}. Finally, we give a separate
discussion of complex M-theory in appendix \ref{M-theory section}.

When preparing this paper we were communicated by Antoine Van
Proeyen that related work in 4d, using the same techniques of
using different reality conditions, is in progress
\cite{Skenderis:2007cx}.

\section{Type II actions}\label{Type II}
In this first section we will show how one can obtain supergravity
actions for different signatures as different real slices of a
single complex action. Sometimes this leads to different
supergravity theories with the same signature.

The starting point of our construction will be a complex action that
then can be reduced to different real actions. In this paper we will
not address the question of how one can in general construct
sensible complex actions or investigate what a general complex
action invariant under some complexified symmetry group looks like.
Instead we will take a more pragmatic approach. The idea is to start
from a known action in terms of some real fields\footnote{By a real
field, we mean a field that satisfies a reality condition, for
instance a Majorana fermion.} that is invariant under some real
symmetry group. The first step is to construct a complexified
version of this action that is invariant under the complexified
symmetry group. We require that the real action we started from can
be obtained from this complexified action by imposing certain
reality conditions and similarly for the symmetries. At this point
one faces the natural question: are there different real slices
leading to other theories? As it will turn out, theories in
different signatures are found by taking different reality
conditions for a single complex action. In the case one has extended
supersymmetry it can even happen that one finds multiple real
theories in one signature. It is these issues that we will work out
in detail for IIA and IIB supergravity in this section.

This general scheme of finding different real actions as consistent
real slices of a given complex action can be applied quite
generally. For the interested reader we have added the same analysis
for M-theory in appendix \ref{M-theory section}. One would expect
the general procedure presented below to hold for all kinds of
theories in various dimensions although subtleties can arise and
some particular details might change from case to case.

\subsection{The complex type II action}\label{Complex Type II}
To start we will deal with the first of the two questions posed
above. We will show how one can find complex actions that can
respectively be restricted to the known actions of IIA and IIB by
reality conditions, and that are furthermore invariant under the
complexified super Poincar\'{e} group. How the different
formulations of the real 10d super Poincar\'{e} algebra can be found
from the unique ten-dimensional complex $\OSp(1|32)$ algebra was
described in detail in \cite{Bergshoeff:2000qu}.

In complexifying an action it is crucial that all fields appear
holomorphically in the complex action. In other words we replace
fields that take values in $\mathbb{R}$ by fields that take values
in $\mathbb{C}$ in such a way that no complex conjugates appear. If
one does the same complexification on the symmetry transformations,
the complexified action is guaranteed to be invariant under these
complex transformations  as checking the invariance is a pure
algebraic computation that nowhere assumes reality of the involved
parameters\footnote{ One might think that complexifying the
supersymmetries in a maximal supergravity theory leads to a
supergravity with 64 supercharges. This is however not the case. One
should view the complexified action as a mathematical tool and not
as a new theory describing new physical degrees of freedom.}.

This procedure of 'holomorphic complexification' is rather
straightforward and only requires some more consideration in case
of the spinors. Usually spinors appear in the action through
bilinears written in terms of the Dirac conjugate
$\dir{\chi}=\chi^\dagger A$. In this form there appears a complex
conjugation and as such the action is not holomorphic in the
spinor $\chi$. There is an easy way around this as using the
reality condition on the spinors the original real action can
equivalently be written in terms of the Majorana conjugate
$\maj{\chi}=\transp{\chi}\calc$. In this form spinors appear
holomorphically and complexification now amounts to ignoring the
reality condition on the spinors.

We will now illustrate this general principle in case of the
ten-dimensional type II theories. For our notations we refer to
appendix \ref{Conventions}.

As a starting point we will take the actions of type IIA and type IIB as given in \cite{Bergshoeff:2001pv}.
These actions have the following field content
\begin{eqnarray}
{\rm IIA}&:&\hskip 1truecm  \left\{
  g_{\mu \nu},
  B_{\mu \nu},
  \phi,
  \C{1}_{\mu},
  \C{3}_{\mu \nu \rho},
  \psi_\mu,
  \lambda
  \right\} \, , \cr
{\rm IIB}&:&\hskip 1truecm  \left\{
  g_{\mu \nu},
  B_{\mu \nu},
  \phi,
  \C{0},
  \C{2}_{\mu \nu},
  \C{4}_{\mu \cdots \rho},
  \psi_\mu,
  \lambda
  \right\} \, .
\label{fcdemo}
\end{eqnarray}
A combined form of the actions is given by (ignoring four fermion
terms)
\begin{eqnarray}
 S &=&
  - \frac{1}{2\kappa_{10}^2}\int d^{10} x \, e
    \Big\{
    e^{-2\phi} \Big[
    -R\big(\omega(e)\big) -4\big( \partial{\phi} \big)^{2}
    +\tfrac{1}{2} H \cdot H  \nonumber\\
 &&  -2\partial^\mu\phi \CHI{1}_{{\mu}}
        + H \cdot \CHI{3}
   +2 \bar{{\psi}}_{{\mu}}{\Gamma}^{{\mu}{\nu}{\rho}}
    {\nabla}_{{\nu}}{\psi}_{{\rho}}
    -2 \bar{{\lambda}}{\Gamma}^{{\mu}}
    {\nabla}_{{\mu}}{\lambda}
    +4 \bar{{\lambda}} {\Gamma}^{{\mu}{\nu}}
    {\nabla}_{{\mu}}{\psi}_{{\nu}}
    \Big]  \label{IIABactiondemo}\\
  && + \sum_{n=0, 1/2}^{3/2, 2}
    \left(\tfrac{1}{2} G^{(2n)} \cdot G^{(2n)}
    +  G^{(2n)} \cdot \PSI{2n}\right) + \tfrac{1}{4} G^{(5)} \cdot
    G^{(5)}+\frac{1}{2} G^{(5)} \cdot \PSI{5}
    - e^{-1}\call_{\text{CS}} \Big\}\,.\nonumber
    \end{eqnarray}
It is understood that the summation in the above action is over
integers ($n=0,1, 2$) in the IIA case and over half-integers
($n=1/2,3/2$) in the IIB case. In the summation range we first write
the lowest value for the IIA case, before the one for the IIB case.
Remember furthermore that $G^{(5)}$ only appears in IIB and
satisfies an additional self-duality constraint $G^{(5)}=\star
G^{(5)}$ that does not follow from the field equations. In the IIA
case, the massive theory contains an additional mass parameter
$G^{(0)}=m$. The Chern-Simons terms are respectively
\begin{eqnarray}
\call_\text{CS}&=&-\varepsilon^{\mu_1\cdots\mu_{10}}\left(\tfrac{1}{4\cdot24^2}
\, \partial_{\mu_1} C^{(3)}_{\mu_2\m_3\m_4} \, \partial_{\mu_5}
C^{(3)}_{\mu_6\m_7\m_8} \, B_{\m_9\m_{10}}
    + \tfrac{1}{2\cdot24^2} \, \Gg{0} \, \partial_{\m_1} \C{3}_{\mu_2\m_3\m_4} \, B^3_{\m_5\ldots\m_{10}}\right.\nonumber\\
    &&+\left.\tfrac{1}{5\cdot16^2} \, \Gg{0}{}^2 \,
    B^5_{\m_1\ldots\m_{10}}\right)
     \qquad\mbox{(IIA)}\label{CS}\,,\\
 \call_\text{CS}   &=-&\frac{1}{3\cdot24^2}\varepsilon^{\mu_1\cdots\mu_{10}}C^{(4)}_{\m_1\mu_2\m_3\m_4}
 \partial_{\mu_5}
 C^{(2)}_{\m_6\m_7}\partial_{\mu_8}B_{\m_9\m_{10}}\qquad\mbox{(IIB)}\,.
\end{eqnarray}
As explained in appendix \ref{spinors}, we work both in IIA and
IIB with an implicit doublet notation for the spinors. The bosonic
fields couple to the fermions via the bilinears $\CHI{1,3}$ and
$\PSI{2n}$, which read
\begin{align}
  \CHI{1}_\mu =
  & -2 \bar{\psi}_\nu \Gamma^\nu \psi_\mu
    -2 \bar{\lambda} \Gamma^\nu \Gamma_\mu \psi_\nu\, , \notag \\
  \CHI{3}_{\mu\nu\rho} =
  & \tfrac{1}{2}\bar{{\psi}}_{{\alpha}}
    {\Gamma}^{[{\alpha}}
    \Gamma_{\mu\nu\rho}
    {\Gamma}^{{\beta}]}
    {\cal P}{\psi}_{{\beta}}
    + \bar{{\lambda}}
    \Gamma_{\mu\nu\rho}{}^{\beta}
    {{\cal P}}{\psi}_{{\beta}}
    -\tfrac{1}{2}\bar{{\lambda}} {\cal P}\Gamma_{\mu\nu\rho}
    {\lambda}\, , \notag \\
  \PSI{2n}_{\mu_1\cdots \mu_{2n}} =
  & {\textstyle\frac{1}{2}}e^{-{\phi}}
    \bar{{\psi}}_{{\alpha}}
    {\Gamma}^{[{\alpha}}
    \Gamma_{\mu_1\cdots \mu_{2n}}
    {\Gamma}^{{\beta}]}
    {\cal P}_n {\psi}_{{\beta}}
    +{\textstyle\frac{1}{2}}e^{-{\phi}}
    \bar{{\lambda}}
    \Gamma_{\mu_1\cdots \mu_{2n}}
    {\Gamma}^{{\beta}}
    {\cal P}_n{\psi}_{{\beta}}\,+ \notag \\
  & -{\textstyle\frac{1}{4}}
    e^{-{\phi}}
    \bar{{\lambda}}
    \Gamma_{ [ \mu_1\cdots \mu_{2n-1}}
    {\cal P}_n \Gamma_{\mu_{2n} ] } {\lambda}\, .
\label{fermionbilinears}
\end{align}
The supersymmetry
rules read (here given modulo cubic fermion terms)
\begin{align}
  \delta_{{\epsilon}} {e}_{{\mu}}{}^{{a}} =
  & \bar{{\epsilon}}{\Gamma}^{{a}} {\psi}_{{\mu}}\, , \notag \\
  \delta_{{\epsilon}} {\psi}_{{\mu}} =
  & \Big( \partial_{{\mu}} +\tfrac{1}{4}
    \not\!{\omega}_{{\mu}}
    +\tfrac{1}{8}{\cal P}\not\!\! {H}_{\mu}
    \Big) {\epsilon}
    +\tfrac{1}{8} e^{{\phi}} \sum_{n=0,1/2}^{3/2,2} \frac{1}{(2n)!}
    \not \! {G}^{(2n)} {\Gamma}_{{\mu}}
    {\cal P}_n{\epsilon} \notag \\ & +\tfrac{1}{16} e^{{\phi}} \frac{1}{5!}
    \not \! {G}^{(5)} {\Gamma}_{{\mu}}
    {\cal P}_{5/2}{\epsilon}\, , \notag \\
  \delta_{{\epsilon}} B_{\mu\nu} =
  & -2 \, \bar{{\epsilon}} \Gamma_{[\mu}
    {\cal P} {\psi}_{\nu]}\, , \notag \\
  \delta_\epsilon \C{2n-1}_{\mu_1\cdots \mu_{2n-1}} =
   & - e^{-\phi} \, \bar{\epsilon} \,
    \Gamma_{[\mu_1\cdots \mu_{2n-2}} \, {\cal P}_n \,
\Big((2n-1) \psi_{\mu_{2n-1}]} - \tfrac{1}{2} \Gamma_{\mu_{2n-1}]}
\lambda\Big)
     \notag \\
    & +(n-1)(2n-1) \, \C{2n-3}_{[\mu_1\cdots \mu_{2n-3}} \,
   \delta_\epsilon B_{\mu_{2n-2}\mu_{2n-1}]}\, , \displaybreak[2] \notag \\
  \delta_{{\epsilon}}{\lambda} =
  & \Big( \! \! \not \! \partial \phi
    + \tfrac{1}{12} \not\!\! {H} {\cal P}\Big) {\epsilon}
    + \tfrac{1}{4} e^{{\phi}} \sum_{n=0,1/2}^{2,5/2} (-)^{2n} \frac{5-2n}{(2n)!}
   \not\! {G}^{(2n)} {\cal P}_n
    {\epsilon}\, , \notag \\
  \delta_{{\epsilon}}{\phi} =
  & \tfrac{1}{2} \, \bar{{\epsilon}}{\lambda}\,.
\label{IIABsusydemo}
\end{align}
Note that for the IIB case $\Gamma _{*}\epsilon =\epsilon$, $\Gamma
_{*}\psi_\m =\psi_\m$ and $\Gamma _{*}\lambda =-\lambda$.

Up till now we have just written down the action of the type IIA/B
in (1,9) signature in a standard form. We will now interpret the
action (\ref{IIABactiondemo}) in a different way, as a complex
action. All fields are now assumed to be complex, both bosonic and
fermionic. For the fermions this means that they are arbitrary
Dirac spinors, as stated before they only appear holomorphically
in the action through their Majorana conjugate
$\maj{\chi}=\transp{\chi}\calc$. The gamma-matrices with flat
indices remain the standard gamma-matrices of (1,9) Minkowski
space. As we now allow the vielbein to be complex, the curved
gamma-matrices will be part of the complexified Clifford algebra,
see appendix \ref{vielbeine} for more details. The supersymmetry
transformations (\ref{IIABsusydemo}) are understood to be complex
in the same way as the action (\ref{IIABactiondemo}). The
complexified action remains invariant under the complexified
supersymmetry transformations as basic manipulations like symmetry
properties of bilinears, gamma-matrix algebra and Fierz identities
are insensitive
 to this complexification. In the same way the complex action is invariant under the complexified Lorentz-group
 $\mathrm{SO}(10,\mathbb{C})$.

\subsection{Back to reality}
Starting from the complex action and supersymmetry transformations
of the previous section we will now explain how one can construct
different real actions by taking different real slices. In this
subsection, we will do a general analysis determining all variant
supergravities. The result is summarized in table
\ref{tbl:theories}. In the next subsection, we will illustrate the
method with some specific examples.

Let us start by explaining what we mean by taking a real slice. A
reality condition on the fields cannot be chosen at will, but has
to satisfy certain consistency conditions. First of all, one can
only impose a limited number of reality conditions on the
fermions. As is explained in appendix \ref{spinors} this leads to
the following general reality conditions on the fermionic fields
(see (\ref{genrealcond10d}))
\begin{eqnarray}
\epsilon^*&=&-\varepsilon\eta^t\a_\epsilon \calc
A\r\epsilon\,, \nonumber \\
\psi_\mu^*&=&-\varepsilon\eta^t\a_\psi \calc
A\r\psi_\m\,, \label{realfermions} \\
\l^*&=&-\varepsilon\eta^t\a_\l \calc A\r\l\,,\nonumber
\end{eqnarray}
where the $\alpha_\chi$ represents a phase factor that can differ
from field to field. On the bosonic fields, a general reality
condition is given by\footnote{To have a uniform notation the
reality condition for $G^{(0)}$ is given in terms of some formal
$C^{(-1)}$. This is just a shorthand implying
$G^{(0)*}=\a_0G^{(0)}$.}:
\begin{eqnarray}
e_{{\mu}}{}^{{a}}{}^*&=&e_{{\mu}}{}^{{a}}\,,\nonumber\\
\f^*&=&\f\,, \nonumber \\
B_{\mu\nu}^*&=&\a_B B_{\mu\nu}\,, \label{realbosons} \\
C^{(2n-1)*}_{\mu_1\cdots \mu_{2n-1}}&=&\a_n C^{(2n-1)}_{\mu_1\cdots
\mu_{2n-1}}\,,\nonumber
\end{eqnarray}
where again the $\alpha$-factors represent phases. Note that we
have already taken the dilaton to be real, as this is the only
condition consistent with reality of the action. We also choose to
work with real vielbeine. This amounts to using the flat
gamma-matrices that are appropriate to a specific signature. The
complex action is written in terms of fixed flat $\Gamma$-matrices
in signature (1,9). In principle one could keep these fixed during
the whole procedure and allow for purely imaginary vielbein
components. Simultaneously redefining the vielbeine and flat
gamma-matrices then brings one back to the case where the
vielbeine are real and the Clifford algebra has the appropriate
signature. For a more thorough and technical discussion of this
point, see appendix \ref{vielbeine}. This reasoning also reveals a
subtlety concerning the Chern-Simons terms. Supersymmetry of the
action (\ref{IIABactiondemo}) is established thanks to the
relation
\begin{equation} \label{defepsilon}
\Gamma_{a_1 \ldots a_n} = - \frac{1}{(10-n)!} \varepsilon_{a_1 \ldots a_{10}} \Gamma_{11} \Gamma^{a_{10} \ldots a_{n+1}} \,.
\end{equation}
This relation is however only valid for the Clifford algebra with signature $(1,9)$. As explained above,
we choose to work
with the Clifford algebra that has the same signature as space-time. For this Clifford algebra,
the relation (\ref{defepsilon})
is changed to
\begin{equation} \label{defepsilonchanged}
\Gamma_{a_1 \ldots a_n} =  \frac{1}{(10-n)!} \varepsilon_{a_1 \ldots a_{10}} i^{t+1} \Gamma_{11} \Gamma^{a_{10} \ldots a_{n+1}} \,.
\end{equation}
Effectively, rewriting (\ref{defepsilonchanged}) to
(\ref{defepsilon}) corresponds to replacing $\varepsilon_{0 \ldots
9}$ by
\begin{eqnarray} \label{refdefeps}
\varepsilon_{0 \ldots 9} &\rightarrow & -(-i)^{t+1} \varepsilon_{0
\ldots 9}\,,\nonumber\\
\varepsilon^{0 \ldots 9} &\rightarrow & -(i)^{t+1} \varepsilon^{0
\ldots 9}\,.
\end{eqnarray}
When going to a real action of a given signature, one has to replace
the $\varepsilon_{0\ldots9}$ in the complex Chern-Simons term via
the above rule to assure invariance under supersymmetry.

The $\alpha$-factors appearing in the reality conditions on the
bosons and the fermions are not independent. Demanding a real
action and consistency with supersymmetry relates them. The latter
means that both sides of the supersymmetry rules should have the
same behaviour under complex conjugation. In this way, the reality
conditions on the fermions determine those of the bosons.
Analyzing this in detail leads to the relations
\begin{eqnarray}
\a_{\epsilon}&=&\a_{\psi}\,,\nonumber\\
\a_\l^2&=&\a_\psi^2=(-\eta)^{t+1}\,,\nonumber\\
\a_\l&=&(-)^{t+1}\eta\transp{\rho}\s\r\s\,\a_\psi\,,\label{alphas}\\
\a_H&=&\transp{\r}\s^{t+1}\s_3\s^{t+1}\r\s_3\,,\nonumber\\
\a_n&=&(-)^{(2n+1)t}(-\eta)^{(2n+1)}\transp{\r}\s^t\calp_{n}\s^{t+1}\r
\calp_{n}^{-1}\s\,.\nonumber
\end{eqnarray}
The possible solutions of these equations lead to consistent
reality conditions on all fields. They are summarized in table \ref{tbl:theories}.
\begin{table}[ht]
\begin{center}
\begin{tabular}{|c|c|cc|c|ccc|c|}
\hline
  &  \multicolumn{4}{c|}{A}& \multicolumn{4}{c|}{B} \\
\hline $t$ mod 4 &0&\multicolumn{2}{c|}{1}& 2&
\multicolumn{3}{c|}{1}&3\\
type& *M$^+$& MW & *MW & M$^+$
& MW & *MW & $'$MW &  SMW\\
$\varepsilon=\eta$&  +& + & + & + & + &  + & + &  +\\
$\r$&  $\s_3$ &$\unity$& $\s_3$ & $\unity$& $\unity$& $\s_3$ & $\s_1$ &  $i\s_2$\\
$\a_\epsilon=\a_\psi$& $i$& $1$& $1$ & $i$&  $1$& $1$ & $1$ & $1$\\
$\a_\l$&  $i$& $1$& $-1$ & $-i$& $1$& $1$ & $1$ & $1$\\
$\a_B$& $-$& + & + & $-$ & + & + &  $-$ &  $-$\\
$\a_{0}=\a_2$, $\a_{1/2}=\a_{5/2}$& $+$&+ & $-$ & $-$ & + & $-$ & $-$ &  $+$\\
$\a_1$, $\a_{3/2}$ & $-$& + & $-$ & $+$ & + & $-$ & $+$  &  $-$\\
$-(i)^{t+1}$&$-i$&1&1&$i$&1&1&1&$-1$\\
\hline
\end{tabular}
  \caption{\sl Possible reality conditions on the fields of type II supergravities. $t$ is the number of time-like
  directions in space-time. The notation concerning the type of fermionic reality condition is explained
  in appendix \ref{spinors}. Every set of reality conditions (column) corresponds to a different
  variant supergravity theory. The last row refers to the additional
  factor for the Chern-Simons terms. From this table the actions and
  supersymmetry transformations of all 10d variant supergravities
  can be constructed.
 }\label{tbl:theories}
\end{center}
\end{table}
Every possible reality condition corresponds to a unique real supergravity theory that has (\ref{IIABactiondemo}) as
complexified action.

Given the data in table \ref{tbl:theories}, the actions and
supersymmetry rules of these variant supergravities can be
explicitly written down. These actions are the complex action
(\ref{IIABactiondemo}), where the fields now obey the reality
properties (\ref{realfermions},\ref{realbosons}), with the
$\alpha$-factors the ones mentioned in table \ref{tbl:theories}. One
notices that in this form some fields might be purely imaginary. In
this case, it is more natural to redefine the fields in terms of
real fields. This leads to a change in sign of e.g. the kinetic
terms of these fields.
In order to write the actions in a more conventional form involving
Dirac conjugates, one can use the following formula equivalent to
(\ref{realfermions}) if (\ref{alphas}) is satisfied:
\begin{equation}
\maj{\chi}=\a_\psi\a_\l\a_{\chi}^*\dir{\chi}\rho\,.
\end{equation}
This allows one to rewrite Majorana conjugates appearing in
(\ref{IIABactiondemo}) in terms of Dirac conjugates. As explained
above in certain signatures one has to multiply the Chern-Simons
term by an additional factor, this factor is given in the last row
of table \ref{tbl:theories}, this same factor also appears in the
(anti) self-duality condition of IIB. The procedure described here
will be illustrated in more detail for some specific examples in
subsection \ref{examples}.

Finally let us give a short overview of the variant theories classified by table \ref{tbl:theories}. Type IIA supergravity
exists in three types of signatures. Note that only in signature $t=1$ mod 4 there are two
different real theories\footnote{The results of table \ref{tbl:theories} almost completely agree with those found in
\cite{Vaula:2002cn} for IIA, with the exception that only in signature (1,9) we find two inequivalent theories. In \cite{Vaula:2002cn}
 additional IIA theories for $t=0$ or 2 mod 4 are presented, which we are not able to reproduce in our framework. }.
 For IIB the situation is similar. Although table \ref{tbl:theories} seems to suggest that there are three different
 theories in (1,9), IIB$^*$ and IIB$'$ are related by a field redefinition that can be interpreted as an S-duality. Note that IIB
 theories only exist in those  signatures where a consistent self-duality condition can be imposed. In our conventions the
 five form is self-dual in signatures with $t=1$ mod 4 and anti self-dual when  $t=3$ mod 4, this is due to the subtleties
  concerning the appearance of $\epsilon_{0\ldots9}$ explained above.

\subsection{Examples}\label{examples}
In this subsection we will illustrate the previously discussed
method of real slices for type II theories in signature (1,9). We
will show how to write down the explicit form of the actions
starting from table \ref{tbl:theories}. To illustrate how to write
down the Chern-Simons terms in case the real slice involves an
additional factor multiplying $\varepsilon_{0\ldots9}$ we discuss
this term in signature (0,10) in detail .

\subsubsection{IIA} Our first example is how one can recover the usual type IIA theory in signature (1,9). The reality conditions
appropriate for this theory are summarized in the second column of
table \ref{tbl:theories}, leading to:
\begin{eqnarray}
\epsilon^*&=&-{\cal C}A \epsilon\,,\nonumber\\
\psi_\mu^*&=&-{\cal C}A \psi_\mu\,,\nonumber\\
\l^*&=&-{\cal C}A \l\label{IIA reality condition A}\\
B_{\mu\nu}^*&=&B_{\mu\nu}\,,\nonumber\\
C^{(2n-1)*}_{\mu_1\cdots \mu_{2n-1}}&=&C^{({2n-1})}_{\mu_1\cdots
\mu_{2n-1}}\,,\,\,\,\,\ (n=0,1,2)\,.\nonumber
\end{eqnarray}
The real action for this theory is the complex action given above
(\ref{IIABactiondemo}) but restricted to the subspace given by these
reality conditions. The Majorana conditions for the spinors
(\ref{IIA reality condition A}) are equivalent to
\begin{eqnarray}
\maj{\epsilon}&=&\dir{\epsilon}\,,\nonumber\\
\maj{\psi}_\mu&=&\dir{\psi}_\mu\label{IIA Majorana conditions}\,,\\
\maj{\l}&=&\dir{\l}\nonumber\,,
\end{eqnarray}
and using these we can write the action (\ref{IIABactiondemo}) in
a standard real form involving Dirac conjugates. Plugging
(\ref{IIA reality condition A}\,-\,\ref{IIA Majorana conditions})
into the action (\ref{IIABactiondemo}) gives
\begin{align}
 S_{\text{IIA}} =
  & - \frac{1}{2\kappa_{10}^2}\int d^{10} x\, e
    \Big\{
    e^{-2\phi} \big[
    -R\big(\omega(e)\big) -4\big( \partial{\phi} \big)^{2}
    +\tfrac{1}{2} H \cdot H
    -2\partial^{{\mu}}{\phi} \chi^{(1)}_{{\mu}}
    + H \cdot \chi^{(3)}\notag \\
  & +2 \dir{{\psi}}_{{\mu}}{\Gamma}^{{\mu}{\nu}{\rho}}
    {\nabla}_{{\nu}}{\psi}_{{\rho}}
    -2 \dir{{\lambda}}{\Gamma}^{{\mu}}
    {\nabla}_{{\mu}}{\lambda}
    +4 \dir{{\lambda}} {\Gamma}^{{\mu}{\nu}}
    {\nabla}_{{\mu}}{\psi}_{{\nu}}
    \big]
    + \sum_{n=0,1,2}
    \tfrac{1}{2} \Gg{2n} \cdot \Gg{2n} \notag \\
  &  + \Gg{2n} \cdot \Psi^{(2n)}
     + e^{-1}
    \varepsilon^{\mu_1\cdots\mu_{10}}\big[\tfrac{1}{4\cdot24^2} \, \partial_{\mu_1} C^{(3)}_{\mu_2\m_3\m_4}
\, \partial_{\mu_5} C^{(3)}_{\mu_6\m_7\m_8} \, B_{\m_9\m_{10}}
    \notag\\
&+ \tfrac{1}{2\cdot24^2} \, \Gg{0} \, \partial_{\m_1} \C{3}_{\mu_2\m_3\m_4} \, B^3_{\m_5\ldots\m_{10}}
   +\tfrac{1}{5\cdot16^2} \, \Gg{0}{}^2 \, B^5_{\m_1\ldots\m_{10}}\big]\Big\}\,,
\label{IIAaction}
\end{align}
where
\begin{eqnarray}
\chi^{(1)}_\mu&=& -2 \dir{\psi}_\nu \Gamma^\nu \psi_\mu
    -2\dir{\lambda} \Gamma^\nu \Gamma_\mu \psi_\nu\, , \nonumber\\
\chi^{(3)}_{\mu\nu\rho}&=&\tfrac{1}{2}\dir{{\psi}}_{{\alpha}}
    {\Gamma}^{[{\alpha}}
    \Gamma_{\mu\nu\rho}
    {\Gamma}^{{\beta}]}\Gamma_{11}
    {\psi}_{{\beta}}
    + \dir{{\lambda}}
    \Gamma_{\mu\nu\rho}{}^{\beta}\Gamma_{11}
    {\psi}_{{\beta}}
    -\tfrac{1}{2}\dir{{\lambda}}\Gamma_{11} \Gamma_{\mu\nu\rho}
    {\lambda}\, , \\
\Psi^{(2n)}_{\mu_1\cdots \mu_{2n}}&=&{\textstyle
\frac{1}{2}}e^{-{\phi}}
    \dir{{\psi}}_{{\alpha}}
    {\Gamma}^{[{\alpha}}
    \Gamma_{\mu_1\cdots \mu_{2n}}
    {\Gamma}^{{\beta}]}
    (\G_{11})^{n} {\psi}_{{\beta}}
    +{\textstyle\frac{1}{2}}e^{-{\phi}}
    \dir{{\lambda}}
    \Gamma_{\mu_1\cdots \mu_{2n}}
    {\Gamma}^{{\beta}}
    (\G_{11})^{n}{\psi}_{{\beta}} \nonumber \\
  && -{\textstyle\frac{1}{4}}
    e^{-{\phi}}
    \dir{{\lambda}}
    \Gamma_{ [ \mu_1\cdots \mu_{2n-1}}
     (\G_{11})^{n}\Gamma_{\mu_{2n} ] } {\lambda}\,\nonumber\,.
\end{eqnarray}
The action (\ref{IIAaction}) is invariant under the following
supersymmetries
\begin{align}
  \delta_{{\epsilon}} {e}_{{\mu}}{}^{{a}} =
  & \dir{{\epsilon}}{\Gamma}^{{a}} {\psi}_{{\mu}}\, , \notag \\
  \delta_{{\epsilon}} {\psi}_{{\mu}} =
  & \Big( \partial_{{\mu}} +\tfrac{1}{4}
    \not\!{\omega}_{{\mu}}
    +\tfrac{1}{8} \Gamma_{11} \not\!\! {H}_{\mu}
    \Big) {\epsilon}
    +\tfrac{1}{8} e^{{\phi}} \sum_{n=0,1,2} \frac{1}{(2n)!}
    \not \! {G}^{(2n)} {\Gamma}_{{\mu}}
    (\Gamma_{11})^n \, {\epsilon}\, , \notag \\
  \delta_{{\epsilon}} B_{\mu\nu} =
  & -2 \, \dir{{\epsilon}} \Gamma_{[\mu}
    \Gamma_{11} {\psi}_{\nu]}\, , \notag \\
  \delta_\epsilon \C{2n-1}_{\mu_1\cdots \mu_{2n-1}} =
   & - e^{-\phi} \, \dir{\epsilon} \,
     \Gamma_{[\mu_1\cdots \mu_{2n-2}} \, (\Gamma_{11})^n \,
     \Big((2n-1) \psi_{\mu_{2n-1}]} - \tfrac{1}{2} \Gamma_{\mu_{2n-1}]}
     \lambda\Big) \notag \\
    & +(n-1)(2n-1) \, \C{2n-3}_{[\mu_1\cdots \mu_{2n-3}} \,
   \delta_\epsilon B_{\mu_{2n-2}\mu_{2n-1}]}\, , \displaybreak[2] \notag \\
  \delta_{{\epsilon}}{\lambda} =
  & \Big( \! \! \not \! \partial \phi
    + \tfrac{1}{12} \not\!\! {H} \Gamma_{11} \Big) {\epsilon}
    + \tfrac{1}{4} e^{{\phi}} \sum_{n=0,1,2} \frac{5-2n}{(2n)!}
   \not\! {G}^{(2n)} (\Gamma_{11})^n
    {\epsilon}\, , \notag \\
  \delta_{{\epsilon}}{\phi} =
  & \tfrac{1}{2} \, \dir{{\epsilon}}{\lambda}\,.
\label{IIAsusy}
\end{align}
As the (1,9) IIA supergravity theory was the theory we started
from before complexifying, taking the real slice was rather
straightforward. Things will become more interesting in case some
fields are purely imaginary. We illustrate this in the following
example.

\subsubsection{IIA$^*$}\label{IIA$^*$ theory}
The action of the IIA* theory in (1,9) can be constructed by using
the third column of table \ref{tbl:theories}, which leads to the
following reality conditions:
\begin{eqnarray}
\epsilon^*&=&-{\cal C} A\G_{11}\epsilon\,,\nonumber\\
\psi_\mu^*&=&-{\cal C}A\G_{11}\psi_\mu\,,\nonumber\\
\l^*&=&{\cal C}A\G_{11}\l\,,\label{IIA* reality condition A}\\
B_{\mu\nu}^*&=&B_{\mu\nu}\,,\nonumber\\
C^{(2n-1)*}_{\mu_1\cdots \mu_{2n-1}}&=&-C^{(2n-1)}_{\mu_1\cdots
\mu_{2n-1}}\,,\,\,\,\,\ (n=0,1,2)\,.\nonumber
\end{eqnarray}
Note that now the reality condition for the Ramond-Ramond fields implies that they are purely
imaginary. It is therefore natural to make a redefinition to real fields. We also prefer to have the same
reality condition for all the fermionic fields. Thus we make the field redefinitions
\begin{eqnarray}
\zeta&=&-i\l\,,\nonumber\\
\A{2n-1}&=&-i\C{2n-1}\,,\label{IIA* field redefinitions}\\
F^{(2n)}&=&-i G^{(2n)}\,.\nonumber
\end{eqnarray}
In this case the relation between Majorana and Dirac conjugate of
the spinors is
\begin{eqnarray}
\maj{\epsilon}&=&-\dir{\epsilon}\G_{11}\,,\nonumber\\
\maj{\psi}_\mu&=&-\dir{\psi}_\mu\G_{11}\,,\label{IIA* Majorana conditions}\\
\maj{\zeta}&=&-\dir{\zeta}\G_{11}\nonumber\,.
\end{eqnarray}
Similarly to the IIA case one can obtain a manifestly real action,
which now reads
\begin{align}
 S_{\text{IIA$^*$}} =
  & - \frac{1}{2\kappa_{10}^2}\int d^{10} x\, e
    \Big\{
    e^{-2\phi} \big[
    -R\big(\omega(e)\big) -4\big( \partial{\phi} \big)^{2}
    +\tfrac{1}{2} H \cdot H
    -2\partial^{{\mu}}{\phi} \xi^{(1)}_{{\mu}}
    + H \cdot \xi^{(3)}+ \notag \\
  & -2 \dir{{\psi}}_{{\mu}}\G_{11}{\Gamma}^{{\mu}{\nu}{\rho}}
    {\nabla}_{{\nu}}{\psi}_{{\rho}}
    -2 \dir{{\zeta}}\G_{11}{\Gamma}^{{\mu}}
    {\nabla}_{{\mu}}{\zeta}
    -4i \dir{{\zeta}}\G_{11} {\Gamma}^{{\mu}{\nu}}
    {\nabla}_{{\mu}}{\psi}_{{\nu}}
    \big]
    - \sum_{n=0,1,2}
    \tfrac{1}{2} F^{2n} \cdot F^{2n} \notag \\
  &  +  F^{2n} \cdot \D^{(2n)}
   -e^{-1} \varepsilon^{\mu_1\cdots\mu_{10}}\big[\tfrac{1}{4\cdot24^2} \, \partial_{\mu_1} A^{(3)}_{\mu_2\m_3\m_4}
\, \partial_{\mu_5} A^{(3)}_{\mu_6\m_7\m_8} \, B_{\m_9\m_{10}}
     \notag\\
&+ \tfrac{1}{2\cdot24^2} \, F^{(0)} \, \partial_{\m_1} A^{(3)}_{\mu_2\m_3\m_4} \, B^3_{\m_5\ldots\m_{10}}
   +\tfrac{1}{5\cdot16^2} \, F^{(0)}{}^2 \, B^5_{\m_1\ldots\m_{10}}\big]\Big\}\,,
\label{IIAstaraction}
\end{align}
where
\begin{eqnarray}
\xi^{(1)}_\mu&=& -2 \dir{\psi}_\nu \Gamma^\nu \G_{11}\psi_\mu
    +2i \dir{\zeta} \Gamma^\nu \Gamma_\mu\G_{11} \psi_\nu\, , \nonumber\\
\xi^{(3)}_{\mu\nu\rho}&=&\tfrac{1}{2}\dir{{\psi}}_{{\alpha}}
    {\Gamma}^{[{\alpha}}
    \Gamma_{\mu\nu\rho}
    {\Gamma}^{{\beta}]}
    {\psi}_{{\beta}}
    - i\dir{{\zeta}}
    \Gamma_{\mu\nu\rho}{}^{\beta}
    {\psi}_{{\beta}}
    -\tfrac{1}{2}\dir{{\zeta}} \Gamma_{\mu\nu\rho}
    {\zeta}\, , \\
\D^{(2n)}_{\mu_1\cdots \mu_{2n}}&=&{\textstyle
\frac{i}{2}}e^{-{\phi}}
    \dir{{\psi}}_{{\alpha}}
    {\Gamma}^{[{\alpha}}
    \Gamma_{\mu_1\cdots \mu_{2n}}
    {\Gamma}^{{\beta}]}
    (\G_{11})^{n+1} {\psi}_{{\beta}}
    +{\textstyle\frac{1}{2}}e^{-{\phi}}
    \dir{{\zeta}}
    \Gamma_{\mu_1\cdots \mu_{2n}}
    {\Gamma}^{{\beta}}
    (\G_{11})^{n+1}{\psi}_{{\beta}} \nonumber \\
  && -{\textstyle\frac{i}{4}}
    e^{-{\phi}}
    \dir{{\zeta}}
    \Gamma_{ [ \mu_1\cdots \mu_{2n-1}}
     (\G_{11})^{n+1}\Gamma_{\mu_{2n} ] } {\zeta}\,.\nonumber
\end{eqnarray}
The action (\ref{IIAstaraction}) is invariant under the
supersymmetries
\begin{align}
  \delta_{{\epsilon}} {e}_{{\mu}}{}^{{a}} =
  & \dir{{\epsilon}}{\Gamma}^{{a}}\G_{11} {\psi}_{{\mu}}\, , \notag \\
  \delta_{{\epsilon}} {\psi}_{{\mu}} =
  & \Big( \partial_{{\mu}} +\tfrac{1}{4}
    \not\!{\omega}_{{\mu}}
    +\tfrac{1}{8} \Gamma_{11} \not\!\! {H}_{\mu}
    \Big) {\epsilon}
    +\tfrac{i}{8} e^{{\phi}} \sum_{n=0,1,2} \frac{1}{(2n)!}
    \not \! {F}^{(2n)} {\Gamma}_{{\mu}}
    (\Gamma_{11})^n \, {\epsilon}\, , \notag \\
  \delta_{{\epsilon}} B_{\mu\nu} =
  & -2 \, \dir{{\epsilon}} \Gamma_{[\mu}
     {\psi}_{\nu]}\, , \notag \\
  \delta_\epsilon A^{(2n-1)}_{\mu_1\cdots \mu_{2n-1}} =
   &  -e^{-\phi} \, \dir{\epsilon} \,
     \Gamma_{[\mu_1\cdots \mu_{2n-2}} \, (\Gamma_{11})^{n+1} \,
     \Big(i(2n-1) \psi_{\mu_{2n-1}]} + \tfrac{1}{2} \Gamma_{\mu_{2n-1}]}
     \zeta\Big)  \notag \\
    & +(n-1)(2n-1) \, A^{(2n-3)}_{[\mu_1\cdots \mu_{2n-3}} \,
   \delta_\epsilon B_{\mu_{2n-2}\mu_{2n-1}]}\, , \displaybreak[2] \notag \\
  \delta_{{\epsilon}}{\zeta} =
  & -i\Big( \! \! \not \! \partial \phi
    + \tfrac{1}{12} \not\!\! {H} \Gamma_{11} \Big) {\epsilon}
    + \tfrac{1}{4} e^{{\phi}} \sum_{n=0,1,2} \frac{5-2n}{(2n)!}
   \not\! {F}^{(2n)} (\Gamma_{11})^n
    {\epsilon}\, , \notag \\
  \delta_{{\epsilon}}{\phi} =
  & -\tfrac{i}{2} \, \dir{{\epsilon}}\G_{11}{\zeta}\,.
\label{IIAstarsusy}
\end{align}
Note that indeed in this real form the Ramond-Ramond fields have
wrong sign kinetic terms. Furthermore, there are additional
factors of $i$ appearing in the supersymmetry transformations with
respect to standard IIA. This is similar to the $i$'s appearing in
the pseudo-supersymmetry of
\cite{Skenderis:2006jq,Skenderis:2006fb}, we will elaborate on
this in section \ref{massiveIIA}. Another difference is the
appearance of the chirality matrix $\G_{11}$ in various spinor
bilinears. They appear for example in the variation of the
dilaton, leading to a different transformation of this field under
parity\footnote{ Under parity we understand the transformation
$x^i\rightarrow -x^i$ that reverses the sign of all 9 space-like
directions.}.

\subsubsection{Chern-Simons terms}
As explained above there are some subtleties concerning the
Chern-Simons terms in certain signatures. Here we will briefly
illustrate how the Chern-Simons term of IIA in (0,10) signature
can be obtained, the other cases proceed analogously. Of the
fields appearing in the IIA Chern-Simons term, $B$ becomes purely
imaginary while the others are real, as can be read from the first
column of table \ref{tbl:theories}. We thus make the redefinition
\begin{equation}
\tilde{B}_{\m\n}=-iB_{\m\n}\,.
\end{equation}
Substituting this in the complex Chern-Simons term (\ref{CS}) and  multiplying with the appropriate
 factor $i$ ( see table \ref{tbl:theories}) gives the following real topological terms:
\begin{eqnarray}
&-\frac{1}{2\kappa_{10}^2}\int\,d^{10}x&\varepsilon^{\mu_1\cdots\mu_{10}}\big[\tfrac{1}{4\cdot24^2}
\, \partial_{\mu_1} C^{(3)}_{\mu_2\m_3\m_4} \, \partial_{\mu_5}
C^{(3)}_{\mu_6\m_7\m_8} \,\tilde B_{\m_9\m_{10}}
    - \tfrac{1}{2\cdot24^2} \, \Gg{0} \, \partial_{\m_1} \C{3}_{\mu_2\m_3\m_4} \, \tilde B^3_{\m_5\ldots\m_{10}}\nonumber\\
   & &+\tfrac{1}{5\cdot16^2} \, \Gg{0}{}^2 \, \tilde B^5_{\m_1\ldots\m_{10}}\big]\,.
\end{eqnarray}
Note that apart from the changes in the Chern-Simons term also the
relation between the real potentials and field strengths gets
modified, e.g.
\begin{equation}
G^{(4)}=dC^{(3)}+d\tilde B\wedge A^{(1)}-G^{(0)}\tilde B^2\,,
\end{equation}
instead of the standard relation (\ref{G2n}).

Similar to this example one can find the Chern-Simons terms and
field strengths in other signatures.

\subsubsection{IIB$^*$}\label{IIB*}
As our final example we derive the action and supersymmetry
equations of IIB$^*$, the alternate real IIB theory in signature
(1,9). The reality conditions are:
\begin{eqnarray}
\epsilon^*&=&{\cal C}A \mathcal{P}\epsilon\,,\nonumber\\
\psi_\mu^*&=&{\cal C}A \mathcal{P}\psi_\mu\,,\nonumber\\
\l^*&=&{\cal C}A \mathcal{P} \l\,,\label{IIB* reality condition}\\
B_{\mu\nu}^*&=&B_{\mu\nu}\,,\nonumber\\
C^{(2n-1)*}_{\mu_1\cdots \mu_{2n-1}}&=&-C^{(2n-1)}_{\mu_1\cdots
\mu_{2n-1}}\,,\,\,\,\,\ (n=1/2,3/2,5/2)\,.\nonumber
\end{eqnarray}
We redefine the imaginary fields in term of real fields as follows:
\begin{equation}\label{RR-fields for IIB^*$}
\begin{aligned}
A^{(2n-1)}& =-i C^{(2n-1)}\,, \\
F^{(2n)}& =-i G^{(2n)}\,.
\end{aligned}
\end{equation}
The reality conditions for the spinors are equivalent to the conditions
\begin{eqnarray}\label{Majorana for IIB^*$}
\maj{\epsilon}&=&\dir{\epsilon} \mathcal{P}\,,\\
\maj{\psi}_\mu&=&\dir{\psi}_\mu \mathcal{P}\nonumber\,,\\
\maj{\l}&=&\dir{\l} \mathcal{P} \nonumber\,.
\end{eqnarray}
Substituting this into the complex IIB action
(\ref{IIABactiondemo}) leads to
\begin{align}
 S_{IIB*} =
  & - \frac{1}{2\kappa_{10}^2}\int d^{10} x e
    \Big\{
    e^{-2\phi} \Big[
    -R\big(\omega(e)\big) -4\big( \partial{\phi} \big)^{2}
    +\tfrac{1}{2} H \cdot H + \nonumber\\
 &    -2\partial^\mu\phi \zeta^{(1)}_{{\mu}}
        + H \cdot {\zeta}^{(3)}
   +2 {\bar{{\psi}}_{{\mu}}}^D\mathcal{P}{\Gamma}^{{\mu}{\nu}{\rho}}
    {\nabla}_{{\nu}}{\psi}_{{\rho}}
    -2 {\bar{{\lambda}}}^D\mathcal{P}{\Gamma}^{{\mu}}
    {\nabla}_{{\mu}}{\lambda}
    +4 {\bar{{\lambda}}}^D\mathcal{P} {\Gamma}^{{\mu}{\nu}}
    {\nabla}_{{\mu}}{\psi}_{{\nu}}
    \Big] \displaybreak[2] \notag \\
  &  - \sum_{n=1/2}^{3/2}
    \left(\frac{1}{2} F^{(2n)} \cdot F^{(2n)}
    +  F^{(2n)} \cdot \Delta^{(2n)}\right) - \frac{1}{4} F^{(5)} \cdot
    F^{(5)}-\frac{1}{2} F^{(5)} \cdot \Delta^{(5)}
    \notag \\
  & - e^{-1}\frac{1}{3\cdot24^2}\varepsilon^{\mu_1\cdots\mu_{10}}A^{(4)}_{\m_1\mu_2\m_3\m_4}
 \partial_{\mu_5} A^{(2)}_{\m_6\m_7}\partial_{\mu_8}B_{\m_9\m_{10}} \Big\}\,.
    \label{IIBstaractiondemo}
\end{align}
This action needs to be supplemented with the usual self-duality for
the RR-five form $F^{(5)}$. The bosonic fields couple to the
fermions via the bilinears
\begin{align}
  \zeta^{(1)}_\mu =
  & -2 {\bar{\psi}_\nu}^D \mathcal{P} \Gamma^\nu \psi_\mu
    -2 \bar{\lambda}^D \mathcal{P} \Gamma^\nu \Gamma_\mu \psi_\nu\, , \notag \\
  \zeta^{(3)}_{\mu\nu\rho} =
  & \tfrac{1}{2}{\bar{{\psi}}_{{\alpha}}}^D
    {\Gamma}^{[{\alpha}}
    \Gamma_{\mu\nu\rho}
    {\Gamma}^{{\beta}]}
    {\psi}_{{\beta}}
    + {\bar{{\lambda}}}^D\
    \Gamma_{\mu\nu\rho}{}^{\beta}
    {\psi}_{{\beta}}
    -\tfrac{1}{2}{\bar{{\lambda}}}^D\Gamma_{\mu\nu\rho}
    {\lambda}\, , \notag \\
  \Delta^{(2n)}_{\mu_1\cdots \mu_{2n}} =-i\Bigl(
  & {\textstyle\frac{1}{2}}e^{-{\phi}}
    {\bar{{\psi}}_{{\alpha}}}^D
    {\Gamma}^{[{\alpha}}
    \Gamma_{\mu_1\cdots \mu_{2n}}
    {\Gamma}^{{\beta}]}
    \mathcal{P}{\cal P}_n {\psi}_{{\beta}}
    +{\textstyle\frac{1}{2}}e^{-{\phi}}
    {\bar{{\lambda}}}^D
    \Gamma_{\mu_1\cdots \mu_{2n}}
    {\Gamma}^{{\beta}}
    \mathcal{P}{\cal P}_n{\psi}_{{\beta}} \notag \\
  & -{\textstyle\frac{1}{4}}
    e^{-{\phi}}
   { \bar{{\lambda}}}^D
    \Gamma_{ [ \mu_1\cdots \mu_{2n-1}}
    \mathcal{P}{\cal P}_n \Gamma_{\mu_{2n} ] } {\lambda}\Bigr)\, .
\label{fermionbilinears}
\end{align}
The supersymmetry rules are
\begin{align}
  \delta_{{\epsilon}} {e}_{{\mu}}{}^{{a}} =
  & {\bar{{\epsilon}}}^D\mathcal{P}{\Gamma}^{{a}} {\psi}_{{\mu}}\, , \notag \\
  \delta_{{\epsilon}} {\psi}_{{\mu}} =
  & \Big( \partial_{{\mu}} +\tfrac{1}{4}
    \not\!{\omega}_{{\mu}}
    +\tfrac{1}{8}{\cal P}\not\!\! {H}_{\mu}
    \Big) {\epsilon}
    +\tfrac{i}{8} e^{{\phi}} \sum_{n=1/2}^{3/2} \frac{1}{(2n)!}
    \not \! {F}^{(2n)} {\Gamma}_{{\mu}}
    {\cal P}_n{\epsilon} \notag \\ & +\tfrac{i}{16} e^{{\phi}} \frac{1}{5!}
    \not \! {F}^{(5)} {\Gamma}_{{\mu}}
    {\cal P}_{5/2}{\epsilon}\, , \notag \\
  \delta_{{\epsilon}} B_{\mu\nu} =
  & -2 \, {\bar{{\epsilon}}}^D  \Gamma_{[\mu}
    {\psi}_{\nu]}\, , \notag \\
  \delta_\epsilon \A{2n-1}_{\mu_1\cdots \mu_{2n-1}} =
   & i e^{-\phi} \, {\bar{\epsilon}}^D \,
    \Gamma_{[\mu_1\cdots \mu_{2n-2}} \, {\cal P}{\cal P}_n \,
\Big((2n-1) \psi_{\mu_{2n-1}]} - \tfrac{1}{2} \Gamma_{\mu_{2n-1}]}
\lambda\Big)
     \notag \\
    & +(n-1)(2n-1) \, \A{2n-3}_{[\mu_1\cdots \mu_{2n-3}} \,
   \delta_\epsilon B_{\mu_{2n-2}\mu_{2n-1}]}\, , \displaybreak[2] \notag \\
  \delta_{{\epsilon}}{\lambda} =
  & \Big( \! \! \not \! \partial \phi
    + \tfrac{1}{12} \not\!\! {H} {\cal P}\Big) {\epsilon}
    + \tfrac{i}{4} e^{{\phi}} \sum_{n=1/2}^{5/2} (-)^{2n} \frac{5-2n}{(2n)!}
   \not\! {F}^{(2n)} {\cal P}_n
    {\epsilon}\, , \notag \\
  \delta_{{\epsilon}}{\phi} =
  & \tfrac{1}{2} \, {\bar{{\epsilon}}}^D{\cal P}{\lambda}\,.
\label{IIBstarsusydemo}
\end{align}
Note that in contrast to the standard IIB action the IIB$^*$
action is no longer invariant under the full S-duality group, but
gets mapped to the IIB$'$ theory. Another viewpoint is thus that
IIB$'$ is nothing else than a field redefinition of IIB$^*$. As
such we will not construct its action and supersymmetry
transformations here. They can be obtained either from performing
an S-duality or taking a real slice with the appropriate reality
conditions in table \ref{tbl:theories}.

\subsection{Extended vs unextended supersymmetry}\label{Truncation to $N=1$ theories}
It might be remarkable that in certain signatures different real
slices exist while in others only one real theory is consistent.
This is related to the number of independent supersymmetries.
Although we always discussed theories with 32 real supercharges,
this does not necessarily mean that their supersymmetry is extended.
Depending from signature to signature the dimension of a real
irreducible spinor is 16 or 32. Only in the signatures in which it
is 16, and thus the 32 supercharges imply extended supersymmetry,
different real slices can occur. This can be understood as in this
case different reality conditions can be imposed on the two
independent 16-dimensional spinors.

This suggests that in any signature only one real slice of the
complex 10d $N=1$ supergravities exists. These $N=1$ supergravities
can be seen as truncations of the type II theories by a
$\mathbb{Z}_2$ truncation. Thus one would expect both the standard
theories and their star versions to truncate to the same theory. We
will now show that this is indeed the case in IIA. The truncation is
made by only keeping the fields invariant under the fermion number
symmetry \cite{Bergshoeff:2001pv}:
\begin{align}
  \big\{ \phi, g_{\mu \nu}, B_{\mu \nu} \big\}
  & \rightarrow \big\{ \phi, g_{\mu \nu}, B_{\mu \nu} \big\}\, , \notag \\
    \big\{\C{2n-1}_{\mu_1 \cdots \mu_{2n-1}} \big\}
  & \rightarrow - \big\{ \C{2n-1}_{\mu_1 \cdots \mu_{2n-1}} \big\} \, ,   \label{Omegahat}\\
    \big\{ \psi_\mu , \lambda, \epsilon \big\}
  & \rightarrow  \G_{11}
    \big\{ \psi_\mu , - \lambda, \epsilon \big\}\, . \notag
\end{align}
One can see that both IIA and IIA$^*$ project to the same theory
under identification by this symmetry as this identification is
equivalent to demanding the reality conditions (\ref{IIA reality
condition A}) and (\ref{IIA* reality condition A}) to be identical.
The other IIA truncation that is given in \cite{Bergshoeff:2001pv}
is no longer consistent.

The situation is similar in IIB. For IIB  in (1,9) signature both
truncations given in \cite{Bergshoeff:2001pv} lead to the same
result, for IIB$^*$ only one truncation is consistent with the
reality properties of the spinors while the other identification
is the only consistent one for IIB$'$. In the end all possible
truncations lead to the same $N=1$ theory.

\section{Domain-walls and cosmologies}\label{pseudo}
In this section we will apply the previously discussed method of
complex actions and real slices to construct and relate different
real solutions. We discuss two examples. Our first example is in
massive IIA, where we find a realisation of the
domain-wall/cosmology correspondence of
\cite{Skenderis:2006fb,Skenderis:2006jq,Skenderis:2006rr} in a
supersymmetric theory. After that we look at 9d gauged maximal
supergravity. This as an example to show that our method works in
different dimensions. Note that our examples are all in theories
with extended supersymmetry, as this seems necessary to be able to
take different real slices in the same signature with our method.

In this section we work in the Einstein frame (E), related to the
string frame (S) via $ g_{\mu\nu}^{(S)}=e^{\phi/2}g_{\mu\nu}^{(E)}$.

\subsection{10d Massive IIA/A$^*$}\label{massiveIIA}
We truncate the complex massive IIA theory (\ref{IIABactiondemo}) to
the following action
\begin{align}
 S_{\text{mIIA}} =
  &  \frac{1}{2\kappa_{10}^2}\int d^{10} x\, e\,\Bigl(
    R -\frac{1}{2}\big( \partial{\phi} \big)^{2}
    -     \tfrac{1}{2} e^{5\phi/2}m^2\Bigr)    \,,
    \label{DW action}
\end{align}
where $m=G^{(0)}$ is the Romans' mass parameter. The fermionic
part of the truncated supersymmetry transformations is
\begin{equation}
\begin{aligned}
   \delta{\psi_{\mu}} =&\Bigl(\nabla_{\mu}-\frac{1}{32} W\Gamma_{\mu}\Bigr)\epsilon\,,\\
  \delta_{{\epsilon}}{\lambda} = &
   \Big( \! \! \not \! \partial \phi
    +    \frac{\delta W}{\delta \phi}\Bigl)
    {\epsilon}\,.
\label{DW SUSY}
\end{aligned}
\end{equation}
For our theory (\ref{DW action}) the scalar potential and
superpotential are respectively
\begin{equation}
V=\frac{1}{2}(\frac{\delta W}{\delta
\phi})^2-\frac{9}{32}W^2=\frac{1}{2}e^{5\phi/2} m^2\,,\,\,\,\,
W=e^{5\phi/4}m\,.
\end{equation}
The complex equations of motion are given by
\begin{equation}
\begin{aligned}
0&=\frac{1}{e}\partial_{\mu}\Bigl(e g^{\mu\nu}\partial_{\nu}\phi\Bigr)-\frac{5}{4}e^{5\phi/2}m^2\,,\\
G_{\mu\nu}&=\frac{1}{2}\Bigl(\partial_{\mu}\phi\partial_{\nu}\phi-\frac{1}{2}g_{\mu\nu}
(g^{\eta\rho}\partial_{\eta}\phi\partial_{\rho}\phi)-\frac{1}{2}g_{\mu\nu}e^{5\phi/2}m^2\Bigr)\,.\label{eomcompdom}
\end{aligned}
\end{equation}
We propose the following complex ansatz for a supersymmetric
solution. As we will show, it can be seen as the complexification of
both a domain-wall and a cosmology:
\begin{eqnarray}
e^0_\mu&=&a_0H^{1/16}\delta_\m^0\,,\nonumber\\
e^i_\mu&=&a_i H^{1/16}\delta_\m^i \qquad (i=1\ldots 8)\,,\nonumber\\
e^9_\mu&=&a_9H^{9/16}\delta_\m^9\,,\\
\phi&=&\frac{-5}{4}\log H\,,\nonumber
\end{eqnarray}
here $a_a$ are some constant complex numbers and $H$ is a complex
function depending only on the coordinate $x^9$. The complex metric
is given by $g_{\mu\nu}=e_{\m}^{\,\,a}
e_{\n}^{\,\,b}\eta_{ab}^{(1,9)}$, as in the previous section
\ref{Type II}. For this ansatz the equations of motion
(\ref{eomcompdom}) and the supersymmetry condition from (\ref{DW
SUSY}) reduce to:
\begin{equation}
\partial_9H=a_9m\,.\label{e.o.m}
\end{equation}
So we find the following complex solution to the complexified
massive theory:
\begin{align}
\rmd {s}^{2} & \, =  H^{1/8}\Bigl(-a_0^2(\rmd x^0)^2 +(a_i)^2(\rmd
x^i)^2\Bigr)
+ H^{9/8} a_9^2(\rmd x^9)^{2}\,,\nonumber\\
e^{{\phi}} & \, =  H^{-5/4}
  \,,  \text{~~~~with~~}
  H =1 + a_9m x^9\,.
\label{complex solution}
\end{align}
It is invariant under the following complex supersymmetries:
\begin{align}
\Gamma_{9} \epsilon=\epsilon\,,\,\,\,\
\epsilon=H^{1/32}\epsilon_0\,,
\end{align}
where $\epsilon_0$ is a constant Dirac spinor.
 In section \ref{Type II} we explained how the complex action (\ref{DW
action}) can give rise to several different real theories by taking
different real slices. If we now apply these reality conditions on
the bosonic fields to our complex solution we will find different
real solutions. The different inequivalent reality properties
consistent with section \ref{Type II} and (\ref{e.o.m}) are given
for some signatures in table \ref{tbl:domainwallreality}, note that
here we allow for imaginary vielbeine. For some comments on how this
relates to section \ref{Type II} we refer to appendix
\ref{vielbeine}.
\begin{table}[ht]
\begin{center}
\begin{tabular}{|c|c|cc|c|}
\hline
 $t$ &0&\multicolumn{2}{c|}{1}& 2\\
 \hline
type & mIIA* &mIIA&mIIA*&mIIA\\
$\a_m$& $+$ & $+$ & $-$ & $-$\\
$\a_\phi$ &$+$&$+$&$+$&$+$\\
$\a^0_\mu$ &$-$&$+$&$-$&$+$\\
$\a^1_\mu$ &$+$&$+$&$+$&$+$\\
$\a^i_\mu$ &$+$&$+$&$+$&$+$\\
$\a^9_\mu$ &$+$&$+$&$-$&$-$\\
$A$ &$\unity$&$\G_0$&$i\G_9$&$i\G_0\G_9$\\ \hline
\end{tabular}
  \caption{\sl Possible reality conditions on the fields of the truncated massive IIA supergravity (mIIA), consistent with the equations of motion.
  $t$ is the number of time-like  directions in space-time. When
  dealing with solutions it is preferable to allow for imaginary
  vielbeine as explained in appendix \ref{vielbeinsolutions}. $A$ is
  the product of all $\G$'s that are time-like in the real theory
  and it appears in e.g. the reality condition for the fermions.
 }\label{tbl:domainwallreality}
\end{center}
\end{table}
Let us illustrate how the complex ansatz reduces to a
supersymmetric domain-wall in massive IIA (mIIA) and a
supersymmetric cosmology in mIIA* by imposing the reality
conditions.

\paragraph{Domain-wall in mIIA}

The standard reality conditions lead to mIIA  and can be found in
the second column of table \ref{tbl:domainwallreality}. As all the
fields are real, the action coincides with (\ref{DW action}). The
complex solution becomes the well known domain-wall or D8-brane of
mIIA:
\begin{align}
d{s}^{2} & \, =  H^{1/8}\Bigl(-(dx^0)^{2} + (\rmd \vec{x})^2\Bigr)
+ H^{9/8} (\rmd x^9)^{2}\,,\nonumber\\
e^{{\phi}} & \, =  H^{-5/4}\,,  \text{~~~~with~~}  H =1 + m x^9\,.
\label{8solution}
\end{align}
The complex supersymmetry variations (\ref{DW SUSY}) become
\begin{equation}
\begin{aligned}
   \delta{\psi_{\mu}} =&\Bigl(\nabla_{\mu}-\frac{1}{32} W\Gamma_{\mu}\Bigr)\epsilon\,,\\
  \delta_{{\epsilon}}{\lambda} = &
   \Big( \! \! \not \! \partial \phi
    +    \frac{\delta W}{\delta \phi}\Bigl)
    {\epsilon}\,,
\label{DW SUSY mIIA}
\end{aligned}
\end{equation}
where $W$ is given by
\begin{equation}
W=e^{5\phi/4}m\,, \,\,\,\, \text{with}\,\, m\,\, \in \,\,
\mathbb{R}\,.
\end{equation}
It is not difficult to verify that the domain-wall (\ref{8solution})
has the following unbroken real supersymmetries
\begin{align}
\Gamma_{9} \epsilon=\epsilon\,,\,\,\,\
\epsilon=H^{1/32}\epsilon_0\,,
\end{align}
where $\epsilon_0$ now is a constant Majorana spinor.

\paragraph{Cosmology in mIIA*}

Alternatively, we can apply the reality conditions of mIIA* to the
complex solution (\ref{complex solution}). As can be read from
table \ref{tbl:domainwallreality} in this case $m$ is purely
imaginary, as are two components of the vielbein: $e_\m{}^0$ and
$e_\m{}^9$. This implies that in this case $a_9=a_0=i$. We
redefine $m=i\tilde m$, $e_\m{}^0=i\tilde e_\m{}^0$,
$e_\m{}^9=i\tilde e_\m{}^9$, $\G^0=i\tilde\G^0$ and
$\G^9=i\tilde\G^9$. Substituting all this in the complex solution
(\ref{complex solution}) gives us a supersymmetric cosmological
solution of mIIA$^*$:
\begin{align}
d{s}^{2} & \, =  H^{1/8}\Bigl((dx^0)^{2} + (\rmd \vec{x})^2\Bigr)
- H^{9/8} (dx^9)^{2}\,,\nonumber\\
e^{{\phi}} & \, =  H^{-5/4}\,,
  \text{~~~~with~~}
  H =1 - \tilde{m} x^9\,,
\label{induced COSMO solution}
\end{align}
where $\tilde{m}=F^{(0)}$. Note that this is the E9-brane of
\cite{Hull:1998vg}. Note that also the real action of mIIA* is
different than that of mIIA. It is given in terms of real fields by
\begin{align}
 S_{\text{mIIA*}} =
  &  \frac{1}{2\kappa_{10}^2}\int d^{10} x\, e\,\Bigl(
    R -\frac{1}{2}\big( \partial{\phi} \big)^{2}
    +     \tfrac{1}{2} e^{5\phi/2}\tilde m^2\Bigr)    \,,
    \label{DW action mIIA*}
\end{align}
with the corresponding supersymmetry variations
\begin{equation}
\begin{aligned}
  \delta{\psi_{\mu}} =&\Bigl(\nabla_{\mu}-\frac{i}{32} \tilde{W}\Gamma_{\mu}\Bigr)\epsilon\,,\\
  \delta{\zeta}=
  &  \Bigl(-i\! \! \not \! \partial \phi
    + \frac{\delta \tilde{W}}{\delta \phi}\Bigr)
    {\epsilon}\,.
\label{COSMO SUSY IIAstar}
\end{aligned}
\end{equation}
The superpotential $\tilde{W}$ is real and given by
\begin{equation}
\tilde{W}=e^{5\phi/4}\tilde{m}\,, \,\,\,\, \text{with}\,\,
\tilde{m}\,\, \in \,\, \mathbb{R}\,.
\end{equation}
As was the case for the domain-wall it is easy to check that the
cosmology (\ref{induced COSMO solution}) preserves the following
supersymmetries
\begin{align}
i\tilde\Gamma_9 \epsilon=\epsilon\,,\,\,\,\
\epsilon=H^{1/32}\epsilon_0\,.
\end{align}
Note that now $\epsilon$ is not a standard Majorana spinor but
satisfies a $^*$MW reality condition instead.

The domain-wall and cosmology presented above are a particular
example of the domain-wall/cosmology correspondence of
\cite{Skenderis:2006fb,Skenderis:2006jq,Skenderis:2006rr}, where an
embedding in extended supergravity is possible. Indeed the truncated
mIIA theory is exactly of the gravity-scalar form as proposed there,
as is its domain-wall solution. Furthermore the truncated mIIA*
theory is equal to the truncated mIIA theory up to a relative sign
in front of the scalar potential. This example places the
domain-wall/cosmology correspondence in a supersymmetric context.
This means that the Killing spinor of the solution generates a
supersymmetry of the theory. Furthermore we see that what was called
a pseudo-Killing spinor in
\cite{Skenderis:2006fb,Skenderis:2006jq,Skenderis:2006rr} now is a
generator of a genuine supersymmetry, but in a star theory. In this
example pseudo-supersymmetry in an extended supersymmetric theory
coincides with the supersymmetry of a superalgebra obeying star
reality conditions.

\subsection{Maximal gauged supergravity in 9d}
In the previous example the only scalar field that played a role
was the dilaton.  In the recent paper \cite{Chemissany:2007fg}
pseudo-supersymmetry was also studied in systems with explicit
multiple scalar fields. In \cite{Sonner:2007cp} it was shown that
for the domain-wall/cosmology correspondence subtleties can appear
when axions are included. In light of this we consider a more
general example with multiple scalar fields including an axion.

The theory we will be working with is given by the following
truncated $N=2$, d=9 massive supergravity Lagrangian
\begin{align}
  \mathcal{L}_{\text{9d}}
    = \tfrac{1}{2}\, e \,
    \bigl [ {R} -
  & \tfrac{1}{2} ( \partial \phi)^2 -\tfrac{1}{2} e^{2\phi}( \partial l)^2
    -\tfrac{1}{2} (\partial \varphi)^2 - V(\phi, l, \varphi)\bigr
    ]\,,
\label{m9DIIBaction}
\end{align}
with $V$ given by
\begin{align}
 V=\frac{1}{2}e^{-2\phi+4\varphi/\sqrt{7}}\Bigl(q_1^2+2e^{2\phi}q_1(-q_2+q_1
 l^2)+e^{4\phi}(q_2+q_1l^2)^2\Bigr)\,.
\label{9d potential}
\end{align}
The details of the reduction from IIB are given in
\cite{Bergshoeff:2002nv,Roest:2004pk}. The scalar fields are given
by $\phi$, $\varphi$ and $l$. The constants $q_1$ and $q_2$
specify the gauging. We group the 9d, 16-component $N=2$ spinors
$\chi_i$ in doublets
\begin{equation}
\chi=\begin{pmatrix}\chi_1\\ \chi_2\end{pmatrix}\,.
\end{equation}
From table \ref{tbl:epsiloneta} we see that $\epsilon=\eta=-1$. In
this section we will use $\mathcal{C}=\gamma_0$. The (1,8)
gamma-matrices $\gamma_{\mu}$ are then purely imaginary. In this
notation the supersymmetry transformations of the fermions are
\begin{eqnarray}
  \delta_{\epsilon} \psi_\mu
    & = &\left[\partial_\mu + \omega_\mu -(\tfrac{1}{4} e^\phi \partial_\mu
    l
    - \tfrac{i}{28} \gamma_\mu W)i\sigma_2\right] \epsilon \,, \nonumber\\
  \delta_{\epsilon} \lambda
    & = & (i\not \! \partial \phi -e^{-\phi}\delta_l W)\sigma_3 \, \epsilon
    +e^\phi (i\not \! \partial l +  e^{-\phi}\delta_\phi W)\sigma_1 \, \epsilon \,,  \\
  \delta_{\epsilon} \tilde\lambda & = & i\not \! \partial \varphi\sigma_3\epsilon + \delta_\varphi W\sigma_1 \,
  \epsilon \,.\nonumber
 \label{Killing-9Dfinal}
\end{eqnarray}
The superpotential $W$ is given by
\begin{align}
W=e^{2\varphi/\sqrt{7}}\Bigl(e^{-\phi}q_1+e^{\phi}(q_2+q_1
l^2)\Bigr)\,. \label{9d superpotential}
\end{align}
The above action and supersymmetry rules can be made complex via
the method of section 2. It turns out that there are two real
slices for signature (1,8), see table \ref{tbl:9d theories}. We
denote the star version of the truncated N=2, d=9 theory by 9d*.
\begin{table}[ht]
\begin{center}
\begin{tabular}{|c|c|c|}
\hline  &9d& 9d* \\
\hline
$\varepsilon=\eta$&  -& - \\
$\r$ &$\unity$&  $\s_3$\\
$\a_{l}$ & + &- \\
$\a_{\epsilon}=\a_{\lambda}=\a_{\tilde{\lambda}}$ & + & +\\
$q_1=q_2$ & + &- \\
 \hline
\end{tabular}
  \caption{\sl The two sets of reality conditions appearing in the truncated N=2, d=9  massive
supergravity Lagrangian leading to signature (1,8).
 }\label{tbl:9d theories}
\end{center}
\end{table}
Inspired by the domain-walls of \cite{Roest:2004pk}, we propose the
following complex ansatz:
\begin{align}
  e^0_\mu&=a_0h^{1/28}\delta_\m^0\,, \notag\\
e^i_\mu&=h^{1/28}\delta_\m^i \qquad (i=1\ldots 7)\,,\notag\\
e^8_\mu&=a_8h^{-3/14}\delta_\m^8\,, \label{9D-DW}\\
  e^\phi & =h^{-1/2} h_1 \,,\,\,\,\ e^{\sqrt{7}\varphi} =
  h^{-1}\,,\,\,\,\
  l = c_1 h_1{}^{-1} \,, \notag\\
    h &= h_1 h_2 - c_1^2 \,,\notag
\end{align}
where $a_a$ and $c_1^2$ are some arbitrary complex constants and
$h_1$ and $h_2$ are functions of $x^8$ only. This ansatz is a
supersymmetric complex solution if
\begin{align}
  \partial_8 h_1 = 2 a_8 q_1  \,, \nonumber\\
   \partial_8 h_2 = 2 a_8 q_2 \,.
\end{align}
In this case it has a complex Killing spinor of the form
\begin{equation}
\epsilon=h^{1/56}(\cos f\unity_2 - i\sin f\,\sigma_2)\epsilon_0\,,
\end{equation}
with
\begin{equation}
f=\frac{1}{4}\arctan\left(\frac{2c_1q_1h^{1/2}}{q_2h_1^2-q_1h+q_1c_1^2}\right)\,,
\end{equation}
 and $\epsilon_0$ is a doublet of constant Dirac
spinors that satisfies
\begin{equation}
\g^8\sigma_2\epsilon_0=\epsilon_0\,.
\end{equation}

As in the previous subsection one can now take two different real
slices leading to a pair of real solutions in signature (1,8).
Taking all the fields in (\ref{9D-DW}) real leads back to the
familiar domain-walls of \cite{Roest:2004pk}:
\begin{align}
  ds^2 & = h^{1/14} (-(dx^0)^2+d\vec{x}^2) + h^{-3/7}(dx^8)^2\, ,\notag \\
  e^\phi & = h^{-1/2} h_1 \,,\quad e^{\sqrt{7}\varphi} = h^{-1}\, , \qquad
  l =  c_1 h_1{}^{-1} \,,\\
  h_1&=2q_1x^8+k_1^2,\ h_2=2q_2x^8+k_2^2\quad\mbox{and}\quad
  h=h_1h_2-c_1^2\,,\notag
\end{align}
where $k_i$ are integration constants. As noted above there is
another set of consistent reality conditions. This second real
slice of (\ref{9D-DW}) will give a cosmological solution. From
table \ref{tbl:9d theories} one can see which fields become purely
imaginary. To write everything in terms of real fields we redefine
$q_i = i \tilde{q}_i$, $c_1= i \tilde{c}_1$ and
$l=i\tilde{l}=i\tilde{c}_1 h_1^{-1}$. Although we did not mention
it in table \ref{tbl:9d theories}, consistency with the equations
of motion also requires $e_\mu{}^0$ and $e_\m{}^8$ to be
imaginary, i.e. $a_8=a_0=i$. The ansatz (\ref{9D-DW}) written in
terms of these real fields gives the following cosmological
solution:
\begin{align}
  ds^2 & = h^{1/14} ((dx^0)^2+d\vec{x}^2) - h^{-3/7}(dx^8)^2\, ,\notag \\
  e^\phi & = h^{-1/2} h_1 \,,\quad e^{\sqrt{7}\varphi} = h^{-1}\, , \qquad
  \tilde{l} =  \tilde{c}_1 h_1{}^{-1} \,,
  \label{9D-DW*}
\end{align}
where
\begin{align}
  h = h_1 h_2 + \tilde{c}_1^2 \,, \qquad
  h_1 = 2  \tilde{q}_1 x^8 +k_1^2 \,, \qquad
  h_2 = 2  \tilde{q}_2 x^8 + k_2^2 \,.
\end{align}
This is a real cosmological solution of the star version of the 9d
theory, of which the action can easily be constructed along the
lines of section \ref{Type II}.

Again we find a natural relation between domain-wall solutions and
cosmological solutions as different real slices of a single
complex solution. In this case the relation between the two real
theories is slightly more involved than just reversing the overall
sign of the scalar potential. It is not difficult to see that the
scalar potential in the 9d* theory is now
\begin{align}
 V=-\frac{1}{2}e^{-2\phi+4\varphi/\sqrt{7}}\Bigl(\tilde q_1^2-2e^{2\phi}\tilde
 q_1(\tilde q_2+\tilde q_1 \tilde l^2)+e^{4\phi}(\tilde q_2-\tilde q_1\tilde l^2)^2\Bigr)\,.
\label{9d potential*}
\end{align}
So apart from an overall change in sign with respect to (\ref{9d
potential}) there are also relative sign changes between the
different terms in the potential. This goes together with a
signature change of the scalar manifold in the 9d* case as the axion
$\tilde l$ has wrong sign kinetic term. Note that if the theory is
truncated by setting the axion to zero we again find an example
where one can embed the correspondence of
\cite{Skenderis:2006fb,Skenderis:2006jq,Skenderis:2006rr} in a
supersymmetric theory. Also in this case the pseudo-supersymmetry of
the cosmology can be interpreted as the vanishing of the fermionic
supersymmetry transformations of a theory with twisted reality
conditions.

\section{Discussion}\label{disc}
In this paper we complexified the type II supergravities and their supersymmetry rules. These complex actions
 do not describe physical theories but are a useful mathematical tool that allows to write down
the actions for all variant supergravities as real slices of the
complex action. We illustrated the method in detail for the
standard type II theories and their corresponding star versions in
signature (1,9). Although we restricted our analysis to 10
dimensions one can generalize it to lower dimensions, for some
related results  see
\cite{Lukierski:1984it,Pilch:1984aw,deWit:1987sn,Park:2001jh,
Behrndt:2003cx,
D'Auria:2002fh,Lu:2003dn,Liu:2003qa,Ferrara:2001sx}. In this paper
we gave an additional example for $N=2$ in 9 dimensions. Note
however that when one continues to lower the dimension, more
possibilities could arise since one can then have extended
supersymmetry with $N>2$. This allows for more general reality
conditions on the fermions than considered in this paper. The
matrix $\rho$ appearing in these reality conditions will in
general be a $N\times N$ matrix. It might be interesting to find
out if for
 $N>2$  there can be more than two inequivalent real slices in certain
 signatures.

In the second part of this paper, we have looked at solutions of
these complex theories and shown that one can obtain solutions of
the different real theories by taking real slices. In particular,
we have seen that in this way supersymmetric domain-walls and
{\text{(pseudo-)}} supersymmetric cosmologies can arise as
different slices of one complex solution. The domain-walls are
solutions in an ordinary supergravity, while the cosmologies arise
as solutions of the star version. In this sense the
pseudo-supersymmetry of cosmologies corresponds to supersymmetry
in the star theory. We presented a ten-dimensional example where
the domain-wall/cosmology correspondence of
\cite{Skenderis:2006fb,Skenderis:2006jq,Skenderis:2006rr} can be
embedded into an extended supergravity context.  In another
example in 9 dimensions we again construct a domain-wall and
corresponding cosmology. A noteworthy feature of this last example
is that the potential no longer gets an overall signflip, but only
certain terms in the potential change sign. Furthermore the scalar
manifold changes signature. This might hint that also in a fake
supergravity context more general changes in the potential could
appear
 under the map of a domain-wall to a cosmology.

\section*{Acknowledgments}
We would like to thank F. Denef, J. De Rydt, G. Gibbons, A.
Keurentjes, L. Martucci, P. Smyth, K. Stelle, L. Tamassia, P.
Townsend, A. Van Proeyen and T. Van Riet for useful discussions. The
authors are supported by the European Commission FP6 program
MRTN-CT-2004-005104 in which E.B. and A.P. are associated to Utrecht
University. The work of A.P. is part of the research programme of
the ``Stichting voor Fundamenteel Onderzoek van de Materie'' (FOM).
J.H. is supported by a Breedte Strategie grant of the University of
Groningen. J.R. and D.V.d.B. are Aspirant FWO Vlaanderen and their
work is supported in part by the FWO Vlaanderen, project G.0235.05
and by the Federal Office for Scientific, Technical and Cultural
Affairs through the 'Interuniversity Attraction Poles Programme -
Belgian Science Policy' P6/11-P. They would like to thank the
Galileo Galilei Institute for Theoretical Physics for its
hospitality and stimulating environment.

\appendix
\section{Conventions}\label{Conventions}
In this paper we work in 'mostly plus' convention, meaning that we
will take  $\eta_{ab}\,=\mathrm{diag}\,(-\ldots-+\ldots+)$, where
the first directions are time-like and the last space-like. We use
Greek indices $\mu,\nu,\rho\ldots$ to denote space-time coordinates
and Latin indices $a,b,c\ldots$ represent tangent directions. They
are related via the vielbein $e_{\mu}^{\,a}$. The determinant of the
vielbein is
denoted by $e$.  
The covariant derivative with respect to general coordinate and
local Lorentz transformations is denoted by $\nabla_\mu$, acting on
tensors $\xi$ and spinors $\chi$ as
\begin{align}
  \nabla_\mu \xi = & \, \partial_\mu \xi\, , \notag \\
  \nabla_\mu \xi^\nu = & \, \partial_\mu \xi^\nu +
\Gamma_{\mu \rho}^{\; \; \; \; \nu} \xi^\rho\, , \notag \\
  \nabla_\mu \chi = & \, \partial_\mu \chi +
\tfrac{1}{4} \omega_{\mu}^{\; \; ab} \Gamma_{ab} \chi\, , \notag \\
  \nabla_\mu \chi^\nu = & \, \partial_\mu \chi^\nu +
\Gamma_{\mu \rho}^{\; \; \; \; \nu} \chi^\rho +\tfrac{1}{4}
\omega_{\mu}^{\; \; ab} \Gamma_{ab} \chi^\n\,.
\end{align}
Here $\Gamma_{\mu \rho}^{\; \; \; \; \nu}$ is the affine connection,
$\omega_{\mu}^{\; \; ab}$ is the spin connection defined by
\begin{equation}
\omega_{\mu\,\,b}^{\,\,a}=e_{\nu}^{\,a}e_{\,b}^{\lambda}\Gamma_{\,\mu\lambda}^{\nu}
-e_{\,b}^{\lambda}\partial_{\mu}e_{\lambda}^{\,a}\,,
\end{equation}
and for the Riemann tensor we use the convention
$\mathcal{R}^{\rho}_{\,\,\mu\nu\sigma}=+\partial_{\nu}\Gamma^{\,\,\,\rho}_{\mu\,\,\sigma}+\ldots$.
Symmetrization and anti-symmetrization are with weight one, slashes
are short notation for $\not\!\!H=H^{\mu \nu \rho }\Gamma _{\mu \nu
\rho }$ and $\not\!\!H_\mu =H_{\mu \nu \rho }\Gamma ^{\nu \rho }$
and the form notations used are
\begin{align}
  P^{(p)} = & \, \frac{1}{p!} P^{(p)}_{\mu_1 \cdots \mu_p} \rmd x^{\mu_1} \wdg \cdots \wdg
     \rmd x^{\mu_p} \,, \notag \\
  P^{(p)}  \cdot Q^{(p)} = & \, \frac{1}{p!}
  P^{(p)}_{\mu_1 \cdots \mu_p} Q^{(p)\,\mu_1 \cdots \mu_p}\,,
  \notag \\
  P^{(p)} \wdg Q^{(q)} =
  & \, \frac{1}{p!q!} P^{(p)}_{\mu_1 \cdots \mu_p}
    Q^{(q)}_{\mu_{p+1} \cdots \mu_{p+q}} \rmd x^{\mu_1} \wdg \cdots \wdg
    \rmd x^{\mu_{p+q}} \,, \notag \\
  P^{(p)n}=&\, P^{(p)}\wdg\ldots\wdg P^{(p)}\qquad(n\mbox{
    times})\,,\notag\\
  \star \, P^{(p)} = & \, \frac{1}{(10-p)!p!}\,e\,
    \varepsilon^{(10)}_{\mu_1 \cdots \mu_{10}} P^{(p)\,\mu_{11-p} \cdots \mu_{10}}
    \rmd x^{\mu_1} \wdg \cdots \wdg \rmd x^{\mu_{10-p}} \,, \notag \\
   \star \star \, P^{(p)} = & \, (-)^{p+1} P^{(p)}\,,\label{formconv}
\end{align}
where  $\varepsilon_{0123\cdots 9}=(-)^t\varepsilon ^{0123\ldots
9}=1$ and $t$ is the number of time-like directions. For notational
convenience we group all potentials and field strengths in the
formal sums
\begin{align}
  {\bf G} = \, \sum_{n=0,1/2}^{2,5/2} G^{(2n)} \,, \hspace{1.5cm}
  {\bf C} = \, \sum_{n=1,1/2}^{2,5/2} \C{2n-1} \,. \label{formsums}
\end{align}
The bosonic field strengths are given by
\begin{align}
  H = d B \,, \qquad
  {\bf G} = d {\bf C} - d B \wdg {\bf C} + G^{(0)} {\bf e}^B \,,
\label{G2n}
\end{align}
where it is understood that each equation involves only one term
from the formal sums (\ref{formsums}) (only the relevant
combinations are extracted). Also we will use the following
abbreviation:
\begin{align}
  {\bf e}^{\pm B} \equiv  \pm B + \tfrac{1}{2} B \wdg B
    \pm \tfrac{1}{3!} B \wdg B \wdg B + \ldots
\end{align}
In writing down type II actions, we use the following definitions
\begin{eqnarray}
{\cal P} &= \Gamma_{11}\otimes\unity_{2}=& \unity_{32} \otimes\
\sigma_{3}\quad {\rm (IIA)} \qquad {\rm or}\qquad
-\unity_{32}\otimes\sigma_3\ {\rm (IIB)}\,,
\end{eqnarray}
and
\begin{equation}
{\cal P}_n = (\Gamma_{11}\otimes\unity_{2})^{n}\\\,\,\,\, {\rm
(IIA)}\,\,\,\,{\rm or}\,\,\,\, \unity_{32}\otimes\sigma^1\ ({\rm
n+1/2\ even}),\ \unity_{32}\otimes i \sigma^2\ ({\rm n+1/2\ odd}) \
\,\,\,\,({\rm IIB})\,.\nonumber
\end{equation}

\section{Spinors and their reality
properties}\label{spinors} In this appendix, we will recall various
properties of Clifford algebras and spinors. The purpose of this
appendix is two-fold. On the one hand it serves to introduce our
conventions and notations. On the other hand, the discussion on the
reality conditions on spinors is also rather crucial for the results
presented in this paper. In the first section of this appendix, we
will recall some general properties of Clifford algebras in various
dimensions and signatures. In the second section, we will then
discuss how appropriate reality conditions can be imposed on the
spinors. The latter discussion will be mainly restricted to 10 and
11 dimensions. A good review concerning the matter presented here is
offered in \cite{VanProeyen:1999ni}, whose conventions we will
mainly follow.

\subsection{Clifford algebras in various dimensions and
signatures}

In this section we will consider arbitrary dimensions $d=t+s$, where
$t$ is the number of time-like and $s$ the number of space-like
directions. The Clifford algebra is then defined by the following
anticommutation relation
\begin{equation} \label{clifford}
\{\Gamma_a, \Gamma_b\} = 2 \eta_{ab} \,,
\end{equation}
where $\eta_{ab} = \mathrm{diag}(-\cdots-+\cdots+)$, writing first
the time-like directions and then the space-like ones.

We will always work with unitary representations of
(\ref{clifford}):
\begin{equation} \label{unitaryrepr}
\Gamma_a^\dag = (-)^t A \Gamma_a A^{-1} \,,
\end{equation}
where we define $A$ to be the product of all time-like
$\Gamma$-matrices : $A = \Gamma_1 \cdots \Gamma_t$. In this way,
time-like $\Gamma$-matrices are anti-hermitian, while the space-like
ones are hermitian.

In even dimensions, we will define the chirality matrix $\Gamma_*$
as follows
\begin{equation} \label{defgamma*}
\Gamma_*  = (-i)^{d/2 + t} \Gamma_1 \cdots \Gamma_d \quad
\Rightarrow \quad (\Gamma_*)^2 = \unity \,.
\end{equation}
When we restrict to 10 dimensions, we will also denote $\Gamma_*$ by
$\Gamma_{11}$. Note that in odd dimensions the product of all
$\Gamma$-matrices is always given by a power of $i$ times the unit
matrix.

One can show that there always exists a unitary matrix
$\mathcal{C}_\eta$ such that
\begin{equation} \label{defCeta}
\transp{\calc}_\eta=-\varepsilon\calc_\eta\qquad\mbox{and}\qquad
\transp{\G}_a=-\eta\calc_\eta\G_a\calc^{-1}_\eta\,,
\end{equation}
where $\varepsilon, \eta$ can be $\pm 1$. In even dimensions, both
signs for $\eta$ are possible, corresponding to the fact that both
$\Gamma_a^T$ and $-\Gamma_a^T$ are representations that are
equivalent to $\Gamma_a$. The two possibilities for the charge
conjugation matrix are then related by
\begin{equation} \label{C+isC-chirop}
\mathcal{C}_{+} = \mathcal{C}_{-} \Gamma_* \,.
\end{equation}
In odd dimensions, due to the constraint on the product of all
$\Gamma$-matrices, only one of the representations $\Gamma_a^T$ or
$-\Gamma_a^T$ is equivalent to $\Gamma_a$ and hence only one sign
for $\eta$ is possible. Once the sign of $\eta$ is fixed, the sign
of $\varepsilon$ can be determined. The possibilities for these
signs are summarized in table \ref{tbl:epsiloneta}.

\begin{table}[ht]
\begin{center}
\begin{tabular}{|c|c|c|c|c|c|c|c|c|}
\hline
  $d$ mod 8 &0&1& 2&3&4&5&6&7\\
\hline $(\epsilon,\eta)$&
$(-,+)$&$(-,-)$&$(-,-)$&$(+,+)$&$(+,+)$&$(+,-)$&$(+,-)$&$(-,+)$\\
&$(-,-)$&&$(+,+)$&&$(+,-)$&&$(-,+)$&\\
\hline
\end{tabular}
  \caption{\sl The possible signs for $\varepsilon$ and $\eta$ for
  all dimensions (modulo 8).
 \label{tbl:epsiloneta}}
\end{center}
\end{table}

Defining the following matrix $B_{\eta}$
\begin{equation} \label{defB}
B_\eta = -\varepsilon \eta^t {\cal C}_\eta A \,,
\end{equation}
equations (\ref{unitaryrepr}) and (\ref{defCeta}) then imply that
\begin{equation} \label{gamma*isequiv}
\Gamma_a^* = (-)^{t+1} \eta B_\eta \Gamma_a B_\eta^{-1} \,.
\end{equation}
As for the ${\cal C}_\eta$-matrix, in even dimensions both signs of
$\eta$ are possible, while in odd dimensions only one possibility
for $\eta$ is allowed. Finally, note that the matrix $B_\eta$
satisfies
\begin{equation} \label{BB*}
B_\eta B_\eta^* = -\varepsilon \eta^t (-)^{t(t+1)/2} \unity \,.
\end{equation}

\subsection{Reality conditions for spinors} \label{realspinors}


In order to describe the reality conditions that can be imposed on
spinors, we will first focus on 11 dimensions, where spinors $\chi$
are 32-component spinors. A general reality condition then has the
form
\begin{equation} \label{genrealcondMtheory}
\chi^* = R \chi \,.
\end{equation}
Consistency with Lorentz transformations then implies that $R =
\alpha_\chi B_\eta$, where $\alpha_\chi$ represents a phase factor.
The most general reality condition that can be imposed in 11
dimensions is then:
\begin{equation} \label{realcondMtheory}
\chi^* = - \varepsilon \eta^t \alpha_\chi {\cal C}_\eta A \chi \,.
\end{equation}
Note that this reality condition can also be stated in terms of
Majorana and Dirac conjugates of the spinors. We will define the
Majorana conjugate $\bar{\chi}$ of a spinor $\chi$ as
\begin{equation} \label{defmajconj}
\bar{\chi} = \chi^T {\cal C}_\eta \,,
\end{equation}
whereas the Dirac conjugate $\bar{\chi}^D$ is given by
\begin{equation} \label{Diracconj}
\bar{\chi}^D = \chi^\dag A \,.
\end{equation}
The reality condition (\ref{realcondMtheory}) then relates these two
conjugates as
\begin{equation} \label{MajisDirac}
\bar{\chi} = \alpha_\chi^{-1} \bar{\chi}^D \,.
\end{equation}
Note that (\ref{realcondMtheory}) does not always define a good
reality condition. Indeed, a consistent reality condition should
obey $\chi^{**} = \chi$, or in other words the matrix $B_\eta
B_\eta^*$ should be equal to the identity. From (\ref{BB*}) one
can infer that this only holds for signatures where $t=1,2 \
\mathrm{mod} \ 4$. These are thus the only signatures where a real
version of 11 dimensional supergravity can be formulated.

In order to formulate reality conditions in 10 dimensions, we will
work with a doublet notation, allowing us to treat type IIA and type
IIB theories in a single framework. The 64-component doublets are
the following
\begin{equation} \label{defdoublet}
\chi = \begin{pmatrix} \chi^+ \\ \chi^- \end{pmatrix} \
(\mathrm{type\ IIA})\,, \qquad \chi = \begin{pmatrix} \chi_1 \\
\chi_2
\end{pmatrix} \ (\mathrm{type\ IIB})\,,
\end{equation}
where $\Gamma_* \chi^{\pm} = \pm \chi^{\pm}$. Gamma-matrices and the
charge conjugation matrix ${\cal C_\eta}$ then act on the doublets
by making the following replacements
\begin{eqnarray} \label{gammaCondoublet}
\Gamma_a & \rightarrow & \Gamma_a \otimes \sigma \,, \\
{\cal C}_\eta & \rightarrow & {\cal C}_\eta \otimes \sigma \,,
\end{eqnarray}
where $\sigma$ is given by $\sigma_1$ in type IIA and by $\unity_2$
in type IIB. Note furthermore that $\Gamma_*$ can be represented by
$\unity_{32} \otimes \sigma_3$ in type IIA and by $\unity_{32}
\otimes \unity_2$ in type IIB. In the following and throughout the
paper we will always assume that matrices act on doublets as
indicated in (\ref{gammaCondoublet}), without writing the tensor
products explicitly.

Using this doublet notation, a general reality condition can now be
denoted as follows:
\begin{equation} \label{genrealcond10d}
\chi^*=-\varepsilon\eta^t\a_\chi \calc_\eta A \r \chi\,,
\end{equation}
where $\alpha_\chi$ again represents a phase factor. The presence of
$-\varepsilon \eta^t {\cal C}_\eta A$ is again dictated by
compatibility with Lorentz transformations. Note that the condition
(\ref{genrealcond10d}) now contains a $2\times 2$-matrix $\rho$,
that can mix the two components of the doublets (\ref{defdoublet});
the action of $\rho$ on a doublet should thus be interpreted as
$\unity_{32} \otimes \rho$. We will take the following possibilities
for $\rho$ :
\begin{equation}
\rho \in \{\unity_2, \sigma_1, \rmi \sigma_2 , \sigma_3 \} \,.
\end{equation}
Note that in the type IIA case the matrix $\rho$ is required to be
diagonal, since complex conjugation should preserve the chirality
of the spinor. In the type IIB case, we do not have to impose this
restriction as both parts of the doublet now have the same
chirality. Note that upon making a field redefinition, the reality
conditions with $\rho = \sigma_1$ and $\rho = \sigma_3$ can be
related \footnote{Explicitly, this redefinition is given by
$\chi_1' = \chi_1 + \chi_2$ and $\chi_2' = \chi_1 - \chi_2$. Note
that this redefinition involves only real numbers. Furthermore as
one can see in table \ref{tbl:theories} in the main text, this
redefinition corresponds to going from IIB$'$ to IIB$^*$.}. We can
thus restrict to $\rho \in {\unity, \rmi \sigma_2, \sigma_3}$
without loss of generality.

Again, the requirement that $\chi^{**} = \chi$ leads to a
non-trivial requirement:
\begin{equation}\label{starisinvol10d}
(\sigma^{t+1}\rho)^2=-\varepsilon\eta^t(-)^\frac{t(t+1)}{2}\,.
\end{equation}
In the IIB case, there is moreover an extra consistency condition,
due to the fact that the theory is chiral. Indeed the reality
condition (\ref{genrealcond10d}) has to respect the chirality, which
in 10 dimensions is only possible when $t$ is odd.

The different reality conditions that can be consistently imposed
are then summarized in table \ref{tbl:realityfordoublets10d}.
\begin{table}[ht]
\begin{center}
\begin{tabular}{|c|cccc|}
\hline
$t$ mod $4$  & 0 & 1  & 2 & 3\\
\hline
 IIA   & \multirow{2}{*}{*M}  & MW      & \multirow{2}{*}{M}     & \multirow{2}{*}{/} \\
     & & *MW& & \\
\hline
 IIB   & \multirow{2}{*}{/} & MW     & \multirow{2}{*}{/}     & \multirow{2}{*}{SMW} \\
     & & *MW & &  \\
\hline
\end{tabular}
  \caption{\sl This table gives all the possible ten-dimensional reality conditions of
  the form (\ref{genrealcond10d}) for a doublet of chiral spinors in type IIA and IIB respectively.
  $t$ denotes the number of time-like dimensions. Here M, *M or SM
respectively stand for $\r=\unity_2$, $\s_3$ or $i\s_2$. The
addition of W means that the reality condition respects chirality of
the spinors.
 }\label{tbl:realityfordoublets10d}
\end{center}
\end{table}
In this table we always choose $\epsilon=\eta=1$. This is possible
as $C_-=C_+\G_{11}$, and thus (\ref{genrealcond10d}) with the choice
$\epsilon=\eta=-1$ can always be rewritten in terms of $C_+$ and
$\eta=\epsilon=1$ by redefining $\rho$ and $\a_\chi$ since $\G_{11}$
can be represented as $\sigma_3$ or $\unity_2$ in IIA respectively
IIB.

Finally a word on notation. Note that in denoting the types of
reality conditions on the fermions in table
\ref{tbl:realityfordoublets10d}, we reserve the * when $\rho =
\sigma_3$ in (\ref{genrealcond10d}). M, MW and SMW then correspond
to what is known in the literature as Majorana, Majorana-Weyl and
symplectic Majorana-Weyl (see for instance
\cite{VanProeyen:1999ni}). Although *M suggests a Majorana
condition, this is not true. For instance, what we have called *M
in Euclidean type IIA, corresponds to what in the literature is
called symplectic Majorana.

\section{Reality of the vielbeine}\label{vielbeine}
\subsection{Imaginary vielbeine and signature change}
In this appendix we will give some more details on the equivalence
between choosing to work with on the one hand fixed flat
gamma-matrices of signature (1,9) and possibly imaginary vielbein
or on the other hand gamma-matrices of the appropriate signature
and a real vielbein.

It is important to stress that the flat gamma-matrices appearing
in the complex action (\ref{IIABactiondemo}) are elements of the
Clifford algebra of signature (1,9) obeying the standard reality
condition\footnote{In this appendix we will make the choice
$\epsilon=\eta=1$.} \begin{equation}
\G^{a*}=-\calc\G_0\G^a\G_0\calc^{-1}. \end{equation} The curved
gamma-matrices $\G_\mu=\G_a e_\m{}^a$ no longer obey a reality
condition as $e_\m{}^a$ (and the other fields) are complex.
Because the vielbein, and thus the metric as well, is complex
there is no longer a concept of space-time signature. Note that
the complex metric is defined as
$g_{\mu\nu}=e_\m{}^ae_\n{}^b\eta^{(1,9)}_{ab}$, where
$\eta^{(1,9)}_{ab}=\mathrm{diag}(-+\ldots+)$.

When we impose reality conditions on the fields appearing in the
action we recover a real theory in a signature that can differ from
the (1,9) signature we started from. This can happen as some
components of the vielbein can be purely imaginary such that
$g_{\m\n}$ is real but has a signature different from that of
$\eta^{(1,9)}_{ab}$.

As explained in the main text a choice of reality conditions for the
fermions determines the reality properties of all the bosonic fields
as well. For the vielbein this happens through the supersymmetry
transformation
\begin{equation}
\delta_{{\epsilon}} {e}_{{\mu}}{}^{{a}} =
  \bar{{\epsilon}}{\Gamma}^{{a}} {\psi}_{{\mu}}\,.\label{evar}
\end{equation}
As explained in appendix \ref{spinors} in the reality conditions
for the spinors (\ref{genrealcond10d}) the operator $A$ appears.
This $A$ is the product of the time-like gamma-matrices. So by
choosing $A$ in the reality conditions for the complex fermions
one decides in which space-time signature the real fermions will
be consistent. As we will see below consistency of the above
supersymmetry variation (\ref{evar}) implies that also the real
metric given by these reality conditions has that signature. If
one for example makes a real slice to a theory in signature
$(t,s)$, $t+s=10$, fermions satisfy the following reality
conditions
\begin{eqnarray}
\epsilon^*&=&-\varepsilon\eta^t\a_\epsilon \calc
A\r\epsilon\,, \nonumber \\
\psi_\mu^*&=&-\varepsilon\eta^t\a_\psi \calc
A\r\psi_\m\,,\label{fermsap}
\end{eqnarray}
with
\begin{equation}
A=(\G_0)(i\G_1)\ldots(i\G_{t-1})\,,\label{A}
\end{equation}
where $\G_a$ are elements of the (1,9) Clifford algebra, i.e.
those appearing in the complex action and (\ref{evar}). Propose
the following reality conditions for the vielbeine:
\begin{equation}
({e}_{{\mu}}{}^{{a}})^* = {\a}_{{\mu}}^{{a}}{e}_{{\mu}}{}^{{a}}\,.
\end{equation}
Using this definition and  (\ref{fermsap}), one can calculate
that\footnote{One has to use that $\a_\epsilon\a_\psi=(-)^{t+1}$,
which follows from analysing the other supersymmetry variations.}
\begin{equation}
{\a}_{{\mu}}^{{a}}=(-)^tA^{-1}\G_0\G^a\G_0A(\G^a)^{-1}\,,\label{realitye}
\end{equation}
by taking the complex conjugate of (\ref{evar}). In the case we take
$A$ of the form (\ref{A}) and divide the index $a$ as $i=1\ldots
t-1$, $j=t\ldots 9$ this implies
\begin{eqnarray}
{\a}_{{\mu}}^{{0}}=1\,, \,\,\,\, {\a}_{{\mu}}^{{i}}= -1\,,
\,\,\,\, {\a}_{{\mu}}^{{j}}=1\,.
\end{eqnarray}
So parts of the vielbein are imaginary and indeed this exactly
implies the metric $g_{\m\n}$ now has the signature $(t,s)$.

Although everything works perfectly in this way it is rather odd to
work with vielbeine that have imaginary components. This is why in
the main text we prefer to work in a formulation where the vielbein
is always completely real. This can be accomplished by
simultaneously redefining the appropriate components
${e}_{{\mu}}^{\,\,{i}}=i{\tilde e}_{{\mu}}^{\,\,{i}}$ and
$\G^i$=$i\tilde\G^i$. It is clear that this redefinition changes the
signature of the flat metric $\eta_{ab}$ as the Clifford algebra now
has signature $(t,s)$. Furthermore everywhere else in the
supersymmetry transformations and the action the vielbeine and
$\G$'s appear in pairs of the form ${e}^{{\mu}}{}_{{a}}\G^a$ or
${e}_{{\mu}}{}^{{a}}\G_a$ and as such always in a combination where
one of the redefined variables appears through its inverse. This
means that we can put tildes everywhere without changing the form of
the expressions or having to add $i$'s or minus signs. One should
read the main text with this redefinition in mind although we did
not explicitly write the tildes, i.e. in section
 \ref{Type II} flat gamma-matrices appearing in real actions and supersymmetry
transformations are always elements of the Clifford algebra that has
the same signature as space-time and all vielbeine are real.

\subsection{Imaginary vielbeine without signature change}\label{vielbeinsolutions}
The discussion above brings about another point. One can also take
some of the components of the vielbein imaginary and still obtain
signature (1,9) for the curved metric. This can be achieved by
taking for instance the following matrix $A$:
\begin{equation}
A=i\G_9\,. \label{strangeA}
\end{equation}
This still leads to a consistent reality condition for the
fermions. As explained before $A$ determines what is space and
what is time in the real slice. The choice (\ref{strangeA})
corresponds to
\begin{eqnarray}
{\a}_{{\mu}}^{{0}}=-1\,, \,\,\,\,{\a}_{{\mu}}^{{i}}= 1\,,\,\,\,\,
{\a}_{{\mu}}^{{9}}=-1\,,\qquad(i=1\ldots 8)\,.
\end{eqnarray}
The naturally redefined $\tilde\eta^{(1,9)}_{ab}$ now has
$\tilde\eta^{(1,9)}_{00}=\tilde\eta^{(1,9)}_{ii}=1$ and
$\tilde\eta^{(1,9)}_{99}=-1$ while the original
$\eta^{(1,9)}_{ab}$ from the complex theory had
$\eta^{(1,9)}_{00}=-1$ and
$\eta^{(1,9)}_{ii}=1=\eta^{(1,9)}_{99}$.

This choice for the vielbein does not lead to new real actions.
Changing the role of different coordinates from time-like to
space-like and vice versa, but keeping the signature fixed,
amounts to no more than a relabelling of the coordinates. The
action and supersymmetry variations are not affected by this
permutation of coordinates.

This is not true however for solutions of its equations of motion.
A generic solution is not invariant under exchange of a time-like
and space-like coordinate. For a complexified version of such a
solution, interchanging coordinates again is equivalent to a
relabelling that does not lead to a different complex solution, as
there is no notion of space or time anymore. So given a real
solution, if we complexify it and then go back to a real form by
imposing different reality conditions it can happen that two
coordinates interchange their space- and time-like character. To
keep track of this effect when taking real slices of a complex
solution it is most practical to work with imaginary vielbeine. In
this way one can see explicitly which coordinates will be
time-like and which space-like in a different real form. One can
see this explicitly at work in section \ref{massiveIIA}.

\section{Complex M-theory}\label{M-theory section}
\subsection{The action}
For completeness we also illustrate our method for M-theory
\cite{Cremmer:1978km}. Since we work in the mostly plus convention,
we use the action as given in \cite{Miemiec:2005ry} (ignoring four
fermion terms)
\begin{eqnarray}\label{The complex M-theory action}
   {\cal S} &=& -\frac{1}{4\kappa_{11}^2}\int d^{11}x \,e \Bigr[ -R
                -2\bar{\psi}_\mu\Gamma^{\mu\nu\rho}
                 \nabla_{\nu} \psi_\rho
                +\frac{1}{2}G^{(4)} \cdot G^{(4)}
                 \nonumber\\
             && +\frac{1}{ 48}
                 \left(
                   \bar{\psi}_\mu\Gamma^{\mu\nu\alpha\beta\gamma\delta}
                   \psi_\nu
                   +12\,\bar{\psi}^\alpha\Gamma^{\gamma\delta}\psi^\beta
                 \right)\,G_{\alpha\beta\gamma\delta}\Bigl]\nonumber\\
             && -\frac{1}{4\kappa_{11}^2}\int d^{11}x \frac{1}{144^2}\,\epsilon^{\alpha_1\ldots\alpha_4
                 \beta_1\ldots\beta_4\mu\nu\rho}G_{\alpha_1\ldots\alpha_4}
                 G_{\beta_1\ldots\beta_4}C_{\mu\nu\rho}\,,
\end{eqnarray}
with the following supersymmetry transformation rules
\begin{eqnarray}
  \delta_{\epsilon}e_\mu{}^a &=&  \bar\epsilon\,\Gamma^a\psi_\mu\,,\nonumber\\
  \delta_{\epsilon}C_{\mu\nu\rho} &=& 3\,\bar\epsilon\,
                               \Gamma_{[\mu\nu}\psi_{\rho]}\,,\nonumber\\
  \delta_{\epsilon}\psi_\mu{} &=& (\partial_{\mu}+\tfrac{1}{4}
    \not\!{\omega}_{{\mu}})\epsilon
                           -\frac{1}{2\cdot 144}
                           \left(
                                 \Gamma^{\alpha\beta\gamma\delta}{}_{\mu}
                                  -8\,\Gamma^{\beta\gamma\delta}
                                     \delta_\mu^\alpha
                           \right)\,\epsilon\,G_{\alpha\beta\gamma\delta}
                           \,,\label{SUSYTRAFO}
\end{eqnarray}
and the definitions
\begin{equation}
\begin{aligned}
      G_{\mu\nu\rho\sigma} &=& 4\,\partial_{[\mu}\,C_{\nu\rho\sigma]}\,.\\
\end{aligned}
\end{equation}
In the above action (\ref{The complex M-theory action}) and
supersymmetry transformations (\ref{SUSYTRAFO}) the spinors appear
through the Majorana conjugate, therefore we can make the theory
complex via the same method as discussed in section \ref{Complex
Type II}. To take real slices we first look at the general 11d
reality conditions
\begin{eqnarray}
\epsilon^*&=&-\varepsilon\eta^t\a_\epsilon \calc
A\epsilon\,,\nonumber\\
\psi_\mu^*&=&-\varepsilon\eta^t\a_\psi \calc
A\psi_\m\,,\nonumber\\
e_{{\mu}}{}^{{a}}{}^*&=&e_{{\mu}}{}^{{a}}\,,\\
G^{*}_{\mu_1\cdots \mu_{4}}&=&\a_G G_{\mu_1\cdots
\mu_{4}}\,.\nonumber
\end{eqnarray}
We choose the vielbein and dilaton real as explained in the main
text. The reality conditions of the spinors determine the reality
conditions of the bosonic fields by the supersymmetry equations
\begin{eqnarray}
\a_{\epsilon}&=&\a_\psi\,,\nonumber\\
\a_{\epsilon}^2&=&(-\eta)^{t+1}\,,\\
\a_G&=&\eta^t \alpha_{\epsilon}^2\,.\nonumber
\end{eqnarray}
The solutions to these equations are summarized in table
\ref{tbl:11d theories}. Note that these classify all possible real
theories coming from the complex action above.
\begin{table}[ht]
\begin{center}
\begin{tabular}{|c|c|c|}
\hline $t$ mod 4 &1&{2} \\\hline
$\varepsilon=\eta$&+ &+\\
$\a_\epsilon=\a_\psi$&  $1$& $i$\\
$\a_G$& + & $-$\\
\hline
\end{tabular}
  \caption{\sl The coefficients given in this table determine
  uniquely all possible real theories in eleven dimensions, how to
  construct their actions and supersymmetry transformations from this table
  is explained in the main text.
 }\label{tbl:11d theories}
\end{center}
\end{table}
Given these consistent reality conditions we can take real slices as
in the type II case.

\subsection{Examples}
\subsubsection{M-theory in (1,10)}
For this signature table \ref{tbl:11d theories} learns us that
\begin{eqnarray}
\epsilon^*&=&-{\cal C}A \epsilon\,,\nonumber\\
\psi_\mu^*&=&-{\cal C}A \psi_\mu\,,\label{M reality
condition}\\
G_{\mu_1\ldots\mu_4}^*&=&G_{\mu_1\ldots\mu_4}\,.\nonumber
\end{eqnarray}
Note that the reality conditions on the spinors can be written as
\begin{eqnarray}
\bar{\epsilon}=\bar{\epsilon}^D\,,\nonumber\\
\bar{\psi_{\mu}}=\bar{\psi_{\mu}}^D\,.
\end{eqnarray}
As all fields are real in this signature, the action and
supersymmetry transformation rules are those of (\ref{The complex
M-theory action}) and (\ref{SUSYTRAFO}) without the need of
redefinitions.

\subsubsection{M-theory in (2,9)}
The reality conditions of table \ref{tbl:11d theories} are in this
case
\begin{eqnarray}
\epsilon^*&=&-i{\cal C}A \epsilon\,,\nonumber\\
\psi_\mu^*&=&-i{\cal C}A \psi_\mu\,,\label{M* reality condition}\\
G_{\mu_1\ldots\mu_4}^*&=&-G_{\mu_1\ldots\mu_4}\,.\nonumber
\end{eqnarray}
Since we prefer real bosonic fields, we introduce
\begin{eqnarray}
A_{\mu_1\ldots\mu_3}=&-iC_{\mu_1\ldots\mu_3}\,,\nonumber\\
F_{\mu_1\ldots\mu_4}=& -i G_{\mu_1\ldots\mu_4}\,.
\end{eqnarray}
For the spinors we also find
\begin{eqnarray}
\bar{\epsilon}=-i\bar{\epsilon}^D\,,\nonumber\\
\bar{\psi_{\mu}}=-i\bar{\psi_{\mu}}^D\,.
\end{eqnarray}
The action of M-theory in (2,9) for these real fields is
\begin{eqnarray}\label{The real M*-theory action}
   {\cal S} &=& -\frac{1}{4\kappa_{11}^2}\int\,e\Bigl[ -R+
                2i\dir{\psi}_\mu\Gamma^{\mu\nu\rho}
                 \nabla_{\nu} \psi_\rho
                -\frac{1}{2} F^{(4)}\cdot F^{(4)}
                 \nonumber\\
             && +\frac{1}{ 48}
                 \left(
                   \dir{\psi}_\mu\Gamma^{\mu\nu\alpha\beta\gamma\delta}
                   \psi_\nu
                   +12\,\dir{\psi^\alpha}\Gamma^{\gamma\delta}\psi^\beta
                 \right)\,F_{\alpha\beta\gamma\delta}\Bigr]\nonumber\\
             && +\frac{1}{4\kappa_{11}^2} \int d^{11}x \,\frac{1}{ 144^2}\epsilon^{\alpha_1\ldots\alpha_4
                 \beta_1\ldots\beta_4\mu\nu\rho}F_{\alpha_1\ldots\alpha_4}
                 F_{\beta_1\ldots\beta_4}A_{\mu\nu\rho}\,,
\end{eqnarray}
and it is invariant under the following supersymmetry rules
\begin{eqnarray}
  \delta_{\epsilon}e_\mu{}^a &=& -i \dir{\epsilon}\,\Gamma^a\psi_\mu\,,\nonumber\\
  \delta_{\epsilon}A_{\mu\nu\rho} &=& -3\,\dir{\epsilon}\,
                               \Gamma_{[\mu\nu}\psi_{\rho]}\,,\nonumber\\
  \delta_{\epsilon}\psi_\mu{} &=& \Bigl(\partial_\mu\,+\,\frac{1}{4}\,
                             \omega_{\mu ab}\,\Gamma^{ab}\Bigr)\epsilon
                           -\frac{i}{2\cdot 144}
                           \left(
                                 \Gamma^{\alpha\beta\gamma\delta}{}_{\mu}
                                  -8\,\Gamma^{\beta\gamma\delta}
                                     \delta_\mu^\alpha
                           \right)\,\epsilon\,F_{\alpha\beta\gamma\delta}
                           \,.\label{M*-theory SUSYTRAFO}
\end{eqnarray}
Note that the Chern-Simons term in (\ref{The real M*-theory
action}) is multiplied by a factor of $i$ coming from the
redefinition of $\epsilon_{0\ldots10}$. For more details about
this procedure see the discussion in the main text.

\providecommand{\href}[2]{#2}\begingroup\raggedright\endgroup

\end{document}